\documentclass[12pt,a4paper,twoside]{article}

\usepackage[dvips]{graphics}
\usepackage{graphicx}
\usepackage{cite}

\def\be{\begin{equation}}
\def\ee{\end{equation}}
\def\bea{\begin{eqnarray}}
\def\eea{\end{eqnarray}}

\newcommand{\bbbar}     {\ensuremath{\mathrm{b\bar{b}}}}

\newcommand{\epem}              {\ensuremath{\mathrm{e^+e^-}}}

\newcommand{\CA}                {\ensuremath{C_{\mathrm{A}}}}
\newcommand{\CF}                {\ensuremath{C_{\mathrm{F}}}}

\newcommand{\nf}                {\ensuremath{n_{\mathrm{f}}}}
\newcommand{\as}                {\ensuremath{\alpha_\mathrm{S}}}
\newcommand{\asi}                {\ensuremath{\alpha_{\mathrm{S},i}}}
\newcommand{\asrs}                {\ensuremath{\alpha_\mathrm{S}(\sqrt{s})}}

\newcommand{\ascu}    {\ensuremath{\alpha_\mathrm{S}^{\mathrm{3}}}}

\newcommand{\asmz}              {\ensuremath{\alpha_\mathrm{S}(M_{\mathrm{Z^0}})}}

\newcommand{\mz}                {\ensuremath{M_{\mathrm{Z^0}}}}

\newcommand{\evis}              {\ensuremath{E_{\mathrm{vis}}}}

\newcommand{\pmiss}     {\ensuremath{p_{\mathrm{miss}}}}
\newcommand{\pbal}              {\ensuremath{p_{\mathrm{bal}}}}

\newcommand{\chisq}     {\ensuremath{\chi^2}}

\newcommand{\xmu}               {\ensuremath{x_{\mu}}}

\newcommand{\ntrkl}             {\ensuremath{N_{\mathrm{long}}}}

\newcommand{\ycut}              {\ensuremath{y_{\mathrm{cut}}}}

\newcommand{\stat}              {\ensuremath{\mathrm{(stat.)}}}

\newcommand{\expt}               {\ensuremath{\mathrm{(exp.)}}}
\newcommand{\had}               {\ensuremath{\mathrm{(had.)}}}
\newcommand{\theo}              {\ensuremath{\mathrm{(theo.)}}}

\newcommand{\rs}                {\ensuremath{\sqrt{s}}}
\newcommand{\rsp}               {\ensuremath{\sqrt{s'}}}

\newcommand{\sigtot} {\ensuremath{\sigma_{\mathrm{tot}}}}

\newcommand{\invpb}     {\ensuremath{\mathrm{pb}^{-1}}}

\newcommand{\py}                {PYTHIA}
\newcommand{\hw}                {HERWIG}
\newcommand{\ar}                {ARIADNE}

\newcommand{\debr}    {DEBRECEN 2.0}
\newcommand{\cdet}    {\ensuremath{C^{\mathrm{detector}}}}
\newcommand{\chad}    {\ensuremath{C^{\mathrm{had}}}}

\newcommand{\result} {
\ensuremath{\asmz=0.1169\pm0.0004\stat\pm0.0012\expt\pm0.0021\had\pm0.0007\theo}}
\newcommand{\restot} {\ensuremath{\asmz=0.1169\pm0.0026~(\mathrm{total~error})}}

\begin{document}

\begin{titlepage}

\centerline{{\large Max-Planck-Institut f\"ur Physik}} \bigskip

\begin{flushright}
 JADE Note 146 \\
 MPP-2004-99 \\
August 6, 2004
\end{flushright}

\bigskip\bigskip\bigskip

\begin{center}
\textbf{
\Large Measurement of the Strong Coupling Constant \boldmath{\as} from the 
Four-Jet Rate in \epem\ Annihilation using JADE data
}
\end{center}

{\Large \par}

\bigskip

\begin{center}
  {\Large J. Schieck, S. Kluth, S. Bethke, P.A. Movilla Fernandez,  C.
    Pahl, and the JADE Collaboration\footnote{See~\cite{naroska87} for
      the full list of authors}}
\end{center}

\par

\bigskip

\begin{abstract}
  Data from \epem\ annihilation into hadrons collected by the JADE
  experiment at centre-of-mass energies between 14~GeV and 44~GeV were
  used to study the four-jet rate as a function of the Durham
  algorithm's resolution parameter \ycut. The four-jet rate was
  compared to a QCD NLO order calculations including NLLA resummation
  of large logarithms.  The strong coupling constant measured from the
  four-jet rate is
\begin{center}
\result, \\
\restot \\
\end{center}
in agreement with the world average.
\end{abstract}

\bigskip\bigskip\bigskip\bigskip\bigskip

{\large \par}

\bigskip\bigskip

\begin{center}{\Large This note describes preliminary JADE results}\end{center}{\LARGE \par}

\vfill

\end{titlepage}

\section{Introduction}

The annihilation of an electron and a positron into hadrons allows
precise tests of Quantum Chromodynamics (QCD).  Multijet rates are
predicted in perturbation theory as functions of the jet-resolution
parameter with one free parameter, the strong coupling constant \as.
Events with four quarks or two quarks and two gluons in the partonic
final state are expected to lead predominantly to events with four
jets in the observed hadronic final state.  Thus, a determination of
the four-jet production rate in hadronic events and fitting the
theoretical prediction to the data provides a means to measure the
strong coupling constant.

Calculations beyond leading order are made possible by theoretical
developments achieved in the last few years.  For multi-jet rates as
well as numerous event shape distributions with perturbative
expansions starting at $\cal{O}(\as)$, matched next-to-leading order
calculations (NLO) and next-to-leading logarithmic approximations
(NLLA) promise precise description of the data over a wide range of
the available kinematic region and centre-of-mass
energy~\cite{durham,dixon97,nagy98a,nagy98b}.

In this analysis we used data collected by the JADE experiment in the
years 1979 to 1986 at the PETRA \epem\ collider at DESY at six
centre-of-mass energies covering the range of 14--44~GeV.  Evidence
for four-jet structure has been reported earlier by
JADE~\cite{bartel82b}.  In a previous OPAL publication a simultaneous
measurement of \as\ and the QCD colour factors with data taken at
$\rs=91$~GeV is described~\cite{OPALPR330}.  The same theoretical
predictions were used for this analysis, with the colour factors \CA\ 
and \CF\ set to the values expected from the QCD SU(3) symmetry group.
A similar analysis was performed by ALEPH using LEP~1 data at
$\rs=91$~GeV~\cite{aleph249}.

The outline of the note is as follows. In section~2, we define the
observable used in the analysis and describe its best available
perturbative predictions. In section~3 the analysis procedure is
explained in detail. Section~4 contains the discussion of the
systematic checks which were performed and the resulting systematic
errors. We collect our results in section~5 and summarize them in
section~6.

\section{Observable}
\label{theory}

Jet algorithms are applied to cluster the large number of particles of
an hadronic event into a small number of jets, reflecting the parton
structure of the event. For this analysis we used the Durham
scheme~\cite{durham}. Defining each particle initially to be a
jet, a resolution variable $y_{ij}$ is calculated for each
pair of jets $i$ and $j$: 
\be
  y_{ij}= \frac{2\mathrm{min}(E_i^2,E_j^2)}{E_{\mathrm{vis}}^2}
          (1-\cos\theta_{ij}),
\ee 
where $E_{i}$ and $E_{j}$ are the energies, $\cos\theta_{ij}$ is the
angle between the two jets and $E_{\mathrm{vis}}$ is the sum of the
energies of all visible particles in the event (or the partons in a
theoretical calculation).  If the smallest value of $y_{ij}$ is less
than a predefined value \ycut, the pair is replaced by a jet with four
momentum $p_{ij}^\mu = p_i^\mu + p_j^\mu$, and the clustering starts
again with $p_{ij}^\mu$ instead of the momenta $p_i^\mu$ and
$p_j^\mu$.  Clustering ends when the smallest value of $y_{ij}$ is
larger than \ycut.  The remaining jets are then counted.

In QCD the fraction of four-jet events $R_4$ is predicted in
NLO as a function of the strong coupling constant \as.
The prediction used here was given by~\cite{nagy98b}:
\be
R_{4}(\ycut)
  = \frac{\sigma_{\mathrm{\mbox{\scriptsize{4-jet}}}}
    (\ycut)}{\sigtot} \\
  = \eta^{2}B_4(\ycut)+\eta^{3}[C_4(\ycut)
    +\frac{3}{2}(\beta_{0}\log{\xmu}-1)\ B_4(\ycut)]
\label{NLOcalc}        
\ee
with \sigtot\ the total hadronic cross-section, $\eta =
\as\CF/(2\pi)$, $\xmu=\mu/\rs$ with $\mu$ the renormalization scale
and \rs\ the centre-of-mass energy, $\beta_{0}=(11-2\nf/3)$ with \nf\ 
the number of active flavours.  The coefficients $B_4$ and $C_4$ were
obtained by integrating the matrix elements for \epem\ 
annihilation into four massless parton final states, calculated by the
program~\debr~\cite{nagy98b}~\footnote{The Durham $y_{\mathrm{cut}}$
  values were chosen to vary in the range between $0.00001$ and
  $0.3162$, similar to the study in~\cite{nagy98b}.}.
Eq.~\ref{NLOcalc} is used to predict the four-jet rate as a function
of \ycut.  The fixed-order perturbative prediction is not reliable for
small values of \ycut, due to terms $\sim\as^{n}\ln^{m}(\ycut)$ that
enhance the higher order corrections.  An all-order resummation, given
in~\cite{durham}, is possible for the Durham clustering algorithm.
The NLLA calculation is combined with the NLO-prediction using the
``modified R-matching'' scheme described in~\cite{nagy98b}.  In
the modified R-matching scheme the terms proportional in $\eta^2$
and $\eta^3$ are removed from the $R^{\mathrm{NLLA}}$ prediction and
the remainder is then added to the $R^{\mathrm{NLO}}$ calculation:
\be
R^{\mathrm{R-match}} 
  = R^{\mathrm{NLLA}}+[\eta^{2}(B_{4}-B^{\mathrm{NLLA}})
    +\eta^{3}(C_{4}-C^{\mathrm{NLLA}}-3/2(B_{4}-B^{\mathrm{NLLA}}))],
\label{NLLA}
\ee
where $B^{\mathrm{NLLA}}$ and $C^{\mathrm{NLLA}}$ are the coefficients
of the expansion of $R^{\mathrm{NLLA}}$ as in Eq.~\ref{NLOcalc}.

\section{Analysis Procedure}

\subsection{The JADE Detector}
\label{sec_detector}

A detailed description of the JADE detector can be found
in~\cite{naroska87}. This analysis relies mainly on the reconstruction
of charged particle trajectories and on the measurement of energy
deposited in the electromagnetic calorimeter.  Tracking of charged
particles was performed with the central detector, which was
positioned in a solenoidal magnet providing an axial magnetic field of
0.48 T. The central detector contained a large volume jet chamber.
Later a vertex chamber close to the interaction point and surrounding
$z$-chambers to measure the $z$-coordinate~\footnote{In the JADE
right-handed coordinate system the $+x$ axis pointed towards the
centre of the PETRA ring, the $y$ axis pointed upwards and the $z$
axis pointed in the direction of the electron beam. The polar angle
$\theta$ and the azimuthal angle $\phi$ were defined with respect to
$z$ and $x$, respectively, while $r$ was the distance from the
$z$-axis.}  were added. Most of the tracking information was obtained
from the jet chamber, which provided up to 48 measured space points
per track, and good tracking efficiency in the region $|\cos
\theta|<0.97$.  Electromagnetic energy was measured by the lead glass
calorimeter surrounding the magnet coil, separated into a barrel
($|\cos \theta|<0.839$) and two end-cap ($0.86<|\cos \theta|<0.97$)
sections.  The electromagnetic calorimeter consisted of 2520 lead
glass blocks with a depth of 12.5 radiation lengths in the barrel
(since 1983 increased to 15.7 in the middle 20\% of the barrel) and
192 lead glass blocks with 9.6 radiation lengths in the end-caps.

\subsection{Data Samples}

The data used in this analysis were collected by JADE between 1979 and
1986 and correspond to a total integrated luminosity of 195 \invpb.
The breakdown of the data samples, mean centre-of-mass energy, energy
range, data taking period, collected integrated luminosities and the size of the
data samples after selection of hadronic events are given in
table~\ref{lumi}.  The data samples were chosen following previous
analyses,
e.g.~\cite{naroska87,jadenewas,OPALPR299,movilla02b,pedrophd}.  The
data are available from two versions of the reconstruction software
from 9/87 and from 5/88.  We used both sets and considered differences
between the results as an experimental systematic uncertainty.

\begin{table}[htb!]
\begin{center}
\begin{tabular}{|r|r|r|r|r|r|} \hline
average       & energy       & year & luminosity  & selected & selected  \\
energy in GeV & range in GeV &      & (\invpb)    & events   & events  \\
 & & &                                            & 9/87     & 5/88 \\
\hline
14.0 & 13.0--15.0 & 1981       & 1.46 & 1722 & 1783 \\
\hline
22.0 & 21.0--23.0 & 1981       & 2.41 & 1383 & 1403 \\
\hline
34.6 & 33.8--36.0 & 1981--1982 & 61.7 & 14213 & 14313 \\
35.0 & 34.0--36.0 & 1986       & 92.3 & 20647 & 20876 \\
\hline
38.3 & 37.3--39.3 & 1985       & 8.28 & 1584 & 1585 \\
\hline
43.8 & 43.4--46.4 & 1984--1985 & 28.8 & 3896 & 4376 \\
\hline
\end{tabular}
\end{center}
\caption{
The average center-of-mass energy, energy range, year
of data taking and integrated luminosity for each data
sample, together with the numbers of selected data events using
the data sample version of 9/87 or 5/88.
}
\label{lumi}
\end{table}

\subsection{Monte Carlo Samples}

Samples of Monte Carlo simulated events were used to correct the data
for experimental acceptance and backgrounds. The process
$\epem\to\mathrm{hadrons}$ was simulated using \py~5.7~\cite{jetset3}.
Corresponding samples using \hw~5.9~\cite{herwig,herwig65} were used
for systematic checks.  The Monte Carlo samples generated at each
energy point were processed through a full simulation of the
JADE detector~\cite{jadesim1,jadesim2,jadesim3}, 
summarized in ~\cite{pedrophd}, and reconstructed in
essentially the same way as the data.
 
 In addition, for comparisons
with the corrected data, and when correcting for the effects of
fragmentation, large samples of Monte Carlo events without detector
simulation were employed, using the parton shower models \py~6.158,
\hw~6.2 and \ar~4.11~\cite{ariadne3}.  All models were adjusted to
LEP~1 data by the OPAL collaboration.

The \ar\ Monte Carlo generator is based on a color dipole mechanism
for the parton shower.  The \py\ and \hw\ Monte Carlo programs use the
leading logarithmic approximation (LLA) approach to model the emission
of gluons in the parton shower.  For the emission of the first hard
gluon, the differences between the LLA approach and a leading order
matrix element calculation are accounted for.  However, the emission
of gluons later on in the parton shower is based only on a LLA cascade
and is expected to differ from a complete matrix element calculation.
For this reason we do expect deviations in the description of the data
by the Monte Carlo models.  Recently new Monte Carlo generators have
been developed, which implement a more complete simulation of the hard
parton emission~\cite{kuhn00}.  However, the models are not yet tuned
to data taken at LEP and were therefore not considered in this
analysis.

\subsection{Selection of Events}

The selection of events for this analysis aims to identify hadronic
event candidates and to reject events with a large amount of energy
emitted by initial state radiation (ISR).  The selection of hadronic
events was based on cuts on event multiplicity (to remove leptonic
final states) and on visible energy and longitudinal momentum balance
(to remove radiative and two-photon events, $\epem \to \epem$
hadrons).  The cuts used are documented 
in~\cite{StdEvSel1,StdEvSel2,StdEvSel3} and summarized in 
a previous publication~\cite{jadenewas}.


Standard criteria were used to select good tracks and clusters of
energy deposits in the calorimeter for subsequent analysis.  
Charged particle tracks were required to have at least 20
hits in r-$\phi$ and at least 12 in r-$z$ in the jet chamber.  The
total momentum was required to be at least 50~MeV.  Furthermore, the
point of closest approach of the track to the collision axis was
required to be less than 5~cm from the nominal collision point in the
$x-y$ plane and less than 35~cm in the $z-$direction. 

In order to mitigate the effects of double counting of energy from
tracks and calorimeter clusters a standard algorithm was
adopted which associated charged particles
with calorimeter clusters, and subtracted the estimated contribution
of the charged particles from the cluster energy. 
Charged particle tracks were assumed to be pions while
the photon hypothesis was assigned to electromagnetic energy clusters.
Clusters in the electromagnetic calorimeter were required to 
have an energy exceeding
0.15~GeV after the subtraction of the expected energy deposit of any
associated tracks.
From all accepted tracks and clusters $i$ the visible energy
$\evis=\sum_i E_i$, momentum balance $\pbal=|\sum_i p_{z,i}|/\evis$ and
missing momentum $\pmiss=|\sum_i \vec{p}_i|$ were calculated.  To
charged particle tracks the pion mass was assigned while the mass
of clusters was assumed to be zero.

Hadronic event candidates were required to pass the following selection 
criteria:
\begin{itemize}
\item The total energy deposited in the electromagnetic calorimeter
  had to exceed 1.2~GeV (0.2~GeV) for $\rs<16$~GeV, 2.0~GeV (0.4~GeV)
  for $16<\rs<24$~GeV and 3.0~GeV (0.4~GeV) for $\rs>24$~GeV in the
  barrel (each endcap) of the detector.
\item The number of good charged particle tracks was required to be
  greater than three reducing $\tau^{+}\tau^{-}$ and two-photon
  backgrounds to a negligible level.
\item For events with exactly four tracks configurations with three
  tracks in one hemisphere and one track in the opposite hemisphere
  were rejected.
\item At least three tracks had to have more than 24 hits in $r-\phi$
  and a momentum larger than 500~MeV; these tracks are called long
  tracks.
\item The visible energy had to fulfill $\evis/\rs>0.5$.
\item The momentum balance had to fulfill $\pbal<0.4$.
\item The missing momentum had to fulfill $\pmiss/\rs<0.3$.
\item The z-coordinate of the reconstructed event vertex had to lie
  within 15~cm of the interaction point.
\item The polar angle of the thrust axis was
  required to satisfy $|\cos(\theta_{\mathrm T})|<0.8$ in order that
  the events be well contained in the detector acceptance.
\end{itemize}
The numbers of selected events for each \rs\ are shown in
table~\ref{lumi} for the two versions of the data.

\subsection{Corrections to the data}
\label{detectorcorrection}

All selected tracks and the electromagnetic calorimeter clusters
remaining after correcting for double counting of energy as described
above were used in the evaluation of the four-jet rate.  The four-jet
rate distribution after all selection cuts had been applied is called
the detector level distribution.

In this analysis events from the process $\epem\rightarrow\bbbar$ were
considered as background, since especially at low \rs\ the large
mass of the b quarks and of the subsequently produced B hadrons will
influence the four-jet rate distribution.  The QCD predictions are
calculated for massless quarks and thus we choose to correct our data
for the presence of \bbbar\ events.

The expected number of \bbbar\ background events $\eta_i$ was
subtracted from the observed number of data events $N_i$ at each
\ycut\ point $i$.  The effects of detector acceptance and resolution
and of residual ISR were then accounted for by a multiplicative
correction procedure.  

Two four-jet rate distributions were formed from Monte Carlo simulated
signal events; the first, at the detector level, treated the Monte
Carlo events identically to the data, while the second, at the
hadron level, was computed using the true momenta of the stable
particles in the event\footnote{ All charged and neutral particles
with a lifetime larger than $3\times 10^{-10}$s were treated as
stable.}, and was restricted to events where $\rsp$, the centre-of-mass
energy reduced due to ISR, satisfied $\rs-\rsp<0.15$~GeV.
The Monte Carlo
ratio of the hadron level to the detector level for each \ycut\ point
$i$, $\cdet_i$, was used as a correction factor for the data.  This
yields finally the corrected number of four jet events at \ycut\ point
$i$ $\tilde{N_{i}}=\cdet_i\cdot(N_{i}-\eta_{i})$.  The hadron level
distribution was then normalized at each \ycut\ point $i$ by
calculating $R_{4,i}=\tilde{N_{i}}/N$, where $N$ is the expected total
number of events.

The detector correction factors \cdet\ as determined using \py\ and
\hw\ are shown in figure~\ref{detcor}.  We observe some disagreement
between the detector corrections calculated using \py\ or \hw\ at low
\rs\ while at larger \rs\ the correction factors agree well within the
regions chosen for comparison with the theory predictions, see below.
The difference in detector corrections was evaluated as an
experimental systematic uncertainty.

A single event will usually contribute to several \ycut\ points in a
four-jet rate distribution and for this reason the data points are
correlated.  The complete covariance matrix $W_{ij}$ was determined
from four-jet rate distributions calculated at the hadron level in the
following way.  Subsamples were built by choosing 1000 events randomly
out of the set of all generated Monte Carlo events. A single event can
show up in several subsamples, but the impact on the final covariance
matrix is expected to be very small and therefore was neglected.  For
every energy point 1000 subsamples were built.  The covariance matrix
was then used to determine the correlation matrix,
$\rho_{ij}=W_{ij}/\tilde{\sigma_{i}}\tilde{\sigma_{j}}$, with
$\tilde{\sigma_{i}}=\sqrt{W_{ii}}$.  

The covariance matrix $V_{ij}$ used in the $\chi^{2}$ fit for the
extraction of \as\ (see section~\ref{fitprocedure} below) was then
determined using the statistical error $\sigma_{i}$ of the data sample
at \ycut\ point $i$ and the correlation matrix $\rho_{ij}$:
$V_{ij}=\rho_{ij}\sigma_{i}\sigma_{j}$.

\section{Systematic Uncertainties}
\label{systematic}

Several sources of possible systematic uncertainties were studied.
All systematic uncertainties were taken as symmetric.  Contributions
to the experimental uncertainties were estimated by repeating the
analysis with varied cuts or procedures.  For each systematic
variation the value of \as\ was determined and then compared to the
result of the standard analysis (default value).  For each variation
the difference with respect to the default value was taken as a
systematic uncertainty.  In the cases of two-sided systematic
variations the larger deviation from the default value was taken as
the systematic uncertainty.
\begin{itemize}
\item In the standard analysis the data version from 9/87 was used.
  As a variation a different data set from 5/88 was used.
\item In the default method the tracks and clusters were
  associated and the estimated energy from the tracks was subtracted.
  As a variation all reconstructed tracks and all electromagnetic
  clusters were used.
\item The thrust axis was required to satisfy $|\cos(\theta_{\mathrm
    T})| < 0.7$.  With this more stringent cut events were restricted
  to the barrel region of the detector, which provides better
  measurements of tracks and clusters compared to the endcaps.
\item Instead of using \py\ for the correction of detector
  effects as described in section~\ref{detectorcorrection}, events
  generated with \hw\ were used.
\item The requirement on missing momentum was dropped or tightened to
  $\pmiss/\rs<0.25$.
\item The requirement on the momentum balance was dropped or tightened
  to $\pbal<0.3$.
\item The requirement on the number of long tracks was tightened to
  $\ntrkl\ge 4$.
\item The requirement on the visible energy was varied to $\evis/\rs>0.45$ and
  $\evis/\rs>0.55$.
\item The amount of subtracted \bbbar\ background was varied
  by $\pm$5\% in order to cover uncertainties in the estimation of
  the background fraction in the data.
\end{itemize}

All contributions listed above were added in quadrature and the result
quoted as the experimental systematic uncertainty.  The dominating
effects were the use of the different data version followed by
employing \hw\ to determine the detector corrections.  In the fits of
the QCD predictions to the data two further systematic uncertainties
were evaluated:

\begin{itemize}
\item The uncertainties associated with the hadronization correction
  (see section~\ref{fitprocedure}) were assessed by using \hw\ and
  \ar\ instead of \py. The larger change in \as\ resulting from these
  alternatives was taken to define the hadronization systematic
  uncertainty.
\item The theoretical uncertainty, associated with missing higher
  order terms in the theoretical prediction, was assessed by varying
  the renormalization scale factor \xmu.  The predictions of an
  all-orders QCD calculation would be independent of \xmu, but a
  finite order calculation such as that used here retains some
  dependence on \xmu.  The renormalization scale \xmu\ was set to 0.5
  and 2.  The larger deviation from the default value was taken as
  theoretical systematic uncertainty.
\end{itemize}

\section{Results}

\subsection{Four-Jet Rate Distributions}

The four-jet rates for the six \rs\ points after subtraction of
background and correction for detector effects are shown in
figures~\ref{hadron} and~\ref{hadron2}.  Superimposed are the
distributions predicted by the \py, \hw\ and \ar\ Monte Carlo models.
In order to make a more clear comparison between data and models, the
inserts in the upper right corner show the differences between data
and each model, divided by the combined statistical and experimental
error at that point. The sum of squares of these differences would, in
the absence of point-to-point correlations, represent a $\chi^{2}$
between data and the model.  However, since correlations are present,
such $\chi^{2}$ values should be regarded only as a rough indication
of the agreement between data and the models. The three models are
seen to describe the data well.

\subsection{Determination of \boldmath{\as}}{
\label{fitprocedure}

Our measurement of the strong coupling constant \as\ is based on
\chisq\ fits of QCD predictions to the corrected four-jet rate
distribution, i.e. the data shown in figures~\ref{hadron}
and~\ref{hadron2}.  The theoretical predictions of the four-jet rate
using the combined $\cal{O}(\ascu)$+NLLA calculation as described in
section~\ref{theory} provide distributions at the parton level.  In
order to confront the theory with the hadron level data, it is
necessary to correct for hadronization effects.  The four-jet rate was
calculated at hadron and parton level using \py\ and, as a
cross-check, with the \hw\ and \ar\ models. The theoretical prediction
is then multiplied by the ratio \chad\ of the hadron and parton level
four-jet rates to correct for hadronization.

The hadronization correction factors \chad\ as obtained from the three
models are shown in figure~\ref{hadcor}.  We find that the
hadronization corrections reach values of down to about 0.5 at low \rs\ 
and there is a significant dependence on \ycut.  For larger \rs\ the
hadronization corrections are closer to one as expected.  We also observe
that the models do not agree well at low \rs.  The differences between
the models will be considered as a systematic uncertainty in the fits.

A $\chi^{2}$-value at each energy point is calculated using the
following formula:
\be
 \chi^{2} = \sum_{i,j}^{n}(R_{4,i}-R(\as)_{4,i}^{\mathrm{theo}})
            (V_{ij})^{-1}(R_{4,j}-R(\as)_{4,j}^{\mathrm{theo}})
\ee
where the indices $i$ and $j$ denote the \ycut\ points in the chosen
fit range and the $R(\as)_{4,i}^{\mathrm{theo}}$ are the predicted
values of the four-jet rate.  The covariance matrix is calculated as
described in section~\ref{detectorcorrection}.  The $\chi^{2}$ value
is minimized with respect to \as\ for each \rs\ separately.
The scale parameter $\xmu$, as discussed in section~\ref{theory}, is
set to 1.  

The fit ranges are shown in table~\ref{fitrange}.  The fit ranges were
determined by requiring that the detector hadronization corrections be
less than $100\%$ in the fit region, and that the $\chi^{2}$/d.o.f.\ 
values changed by less than 2 units when one \ycut\ point is added to
or removed from the fit range.  The fit ranges cover the decreasing
parts of the distributions at large \ycut, where the perturbative QCD
predictions are able to adequately describe the data corrected for
hadronization.  The increasing parts of the distributions at low \ycut\
are dominated by events with more than four jets which cannot be
described accurately by the predictions.  In addition experimental and
hadronization corrections become large in this region.

In figures~\ref{fit_plot} and~\ref{fit_plot2} the hadron level
four-jet distributions for the six energy points are shown together
with the fit result.  The numerical results of the fits are summarized
in table~\ref{fitresults}.  The statistical error corresponds to the
error from the $\chi^{2}$ minimization.  The systematic errors were
determined as described in section~\ref{systematic}.

It is also of interest to combine the measurements of \as\ from the
different centre-of-mass energy points in order to determine a single
value.  This problem has been subject of extensive study by the LEP
QCD working group~\cite{LEPQCDWG}, and we adopt their procedure here.

In brief the method is as follows. The set of \as\ measurements to be
combined are first evolved to a common scale, $Q=\mz$, assuming the
validity of QCD. The measurements are then combined in a weighted
mean, to minimize the $\chi^{2}$ between the combined values and the
measurements. If the measured values evolved to $Q=\mz$ are denoted
$\asi$, with covariance matrix $V^{\prime}$, the combined values,
\asmz, are given by
\be \asmz=\sum w_{i} \asi \;\;\;\; \mathrm{where}
\;\;\;\;
w_{i}=\frac{\sum_{j}(V^{\prime~-1})_{ij}}{\sum_{j,k}(V^{\prime~-1})_{jk}},
\ee
where $i$ and $j$ denote the six individual results.  The difficulty
resides in making a reliable estimate of $V^{\prime}$ in the presence
of dominant and highly correlated systematic errors.  Small
uncertainties in the estimation of these correlations can cause
undesirable features such as negative weights.  For this reason only
experimental systematic errors assumed to be partially correlated
between measurements were taken to contribute to the off-diagonal
elements of the covariance matrix:
$V^{\prime}_{ij}=\min(\sigma^2_{exp,i},\sigma^2_{exp,j})$.  All error
contributions (statistical, experimental, hadronization and scale
uncertainty) were taken to contribute to the diagonal elements.  The
hadronization and scale uncertainties were computed by combining the
\as\ values obtained with the alternative hadronization models, and
from the upper and lower theoretical errors, using the weights derived
from the covariance matrix $V^{\prime}$.  

We find that the fit result from the 14~GeV data has large
hadronization and experimental uncertainties because the corresponding
corrections are large and not well known at this energy.  We therefore
choose to not include this result in the combination.  The result of
the combination using all results with $\rs\ge 22$~GeV is
\begin{displaymath}
 \result\;,
\end{displaymath}
consistent with the world average value of $\asmz=0.1182\pm0.0027$
~\cite{bethke04}.  The weights were 0.18 for 22~GeV, 0.29 for
34.6~GeV, 0.25 for 35~GeV, 0.07 for 38.3~GeV and 0.21 for 44~GeV.  
The results at each energy point are shown in figure~\ref{alphas_fit}
and compared with the predicted running of \as\ based on the world
average value.  For clarity the values from $\rs=34.6$ and 35.0~GeV
have been combined at their luminosity weighted average energy
$\rs=34.8$~GeV using the combination procedure described above.  The
result of ALEPH~\cite{aleph249} and preliminary data from
OPAL~\cite{OPALPN527} are shown as well.

\section{Summary}

In this note we present preliminary measurements of the strong
coupling from the four-jet rate at centre-of-mass energies between 14
and 44~GeV using data of the JADE experiment.  The predictions of the
\py, \hw\ and \ar\ Monte Carlo models tuned by OPAL to LEP~1 data are
found to be in agreement with the measured distributions.

From a fit of QCD NLO predictions combined with resummed NLLA
calculations to the four-jet rate corrected for experimental and
hadronization effects we have determined the strong coupling \as.  The
value of \asmz\ is determined to be \restot.


\begin{thebibliography}{10}

\bibitem{naroska87}
B.~Naroska: Phys. Rep. {\bf 148} (1987) 67

\bibitem{durham}
S.~Catani et~al.: Phys. Lett. B {\bf 269} (1991) 432

\bibitem{dixon97}
L.J. Dixon, A.~Signer: Phys. Rev. D {\bf 56} (1997) 4031

\bibitem{nagy98a}
Z.~Nagy, Z.~Trocsanyi: Phys. Rev. D {\bf 57} (1998) 5793

\bibitem{nagy98b}
Z.~Nagy, Z.~Trocsanyi: Phys. Rev. D {\bf 59} (1998) 014020,\\
  Erratum-ibid.D62:099902,2000

\bibitem{bartel82b}
JADE Coll., W.~Bartel et~al.: Phys. Lett. B {\bf 115} (1982) 338

\bibitem{OPALPR330}
OPAL Coll., G.~Abbiendi et~al.: Eur. Phys. J. C {\bf 20} (2001) 601

\bibitem{aleph249}
ALEPH Coll., A.~Heister et~al.: Eur. Phys. J. C {\bf 27} (2003) 1

\bibitem{jadenewas}
JADE Coll., P.A. {Movilla Fern{\'a}ndez}, O.~Biebel, S.~Bethke, S.~Kluth,
  P.~Pfeifenschneider et~al.: Eur. Phys. J. C {\bf 1} (1998) 461

\bibitem{OPALPR299}
JADE and OPAL Coll., P.~Pfeifenschneider et~al.: Eur. Phys. J. C {\bf 17}
  (2000) 19

\bibitem{movilla02b}
P.A. {Movilla Fern\'andez}: In: {ICHEP 2002: Parallel Sessions}, S.~Bentvelsen,
  P.~de~Jong, J.~Koch, E.~Laenen (eds.), 361. North-Holland, 2003

\bibitem{pedrophd}
P.A. {Movilla Fern\'andez}: {Ph.D.} thesis, RWTH Aachen, 2003, PITHA 03/01

\bibitem{jetset3}
T.~Sj{\"o}strand: Comput. Phys. Commun. {\bf 82} (1994) 74

\bibitem{herwig}
G.~Marchesini et~al.: Comput. Phys. Commun. {\bf 67} (1992) 465

\bibitem{herwig65}
G.~Corcella et~al.: J. High Energy Phys. {\bf 01} (2001) 010


\bibitem{StdEvSel1}
JADE Coll., W.~Bartel et~al.: Phys. Lee {\bf B88} (1979), 171. 
\bibitem{StdEvSel2}
JADE Coll., W.~Bartel et~al.: Phys. Lee {\bf B129} (1983), 145. 
\bibitem{StdEvSel3}
JADE Coll., S.~Bethke et~al.: Phys. Lee {\bf B213} (1988), 235. 


\bibitem{ariadne3}
L.~L{\"o}nnblad: Comput. Phys. Commun. {\bf 71} (1992) 15

\bibitem{kuhn00}
R.~Kuhn, F.~Krauss, B.~Ivanyi, G.~Soff: Comput. Phys. Commun. {\bf 134} (2001)
  223

\bibitem{jadesim1}
E.~Elsen: {\em Detector Monte Carlos}, JADE Computer Note 54.
\bibitem{jadesim2}
E.~Elsen: {\em Multihadronerzeugung in \epem\ Vernichtung beu PETRA-Energien
und Vergleich mit Aussagen der Quantenchromodynamic}, PhD thesis, Universit\"at Hamburg, 1981.
\bibitem{jadesim3}
C.~Bowdery and J.~Olsen: {\em The JADE SUPERVISOR Program}, JADE Computer Note 73.



\bibitem{LEPQCDWG}
{The LEP Experiments (ALEPH, DELPHI, L3 and OPAL) and the LEP QCD Working
  Group}: paper in preparation

\bibitem{bethke04}
S.~Bethke: MPP-2004-88, hep-ex/0407021 (2004)

\bibitem{OPALPN527}
OPAL Coll., G.~Abbiendi et~al.: OPAL physics note PN527 (2004), unpublished

\end{thebibliography}

\clearpage

\section*{ Tables }

\begin{table}[htb!]
\begin{center}
\begin{tabular}{|r|c|} \hline
\rs\ [GeV] & fit range \\
\hline
14.0 & 0.00875 -- 0.02765 \\
22.0 & 0.0049 -- 0.01555 \\
34.6 & 0.0037 -- 0.02765 \\
35.0 & 0.0037 -- 0.02765 \\
38.3 & 0.0021 -- 0.02765 \\
43.8 & 0.0021 -- 0.02075 \\
\hline
\end{tabular}
\end{center}
\caption{ Fit ranges for all energy points }
\label{fitrange}
\end{table}

\begin{table}[htb!]
\begin{center}
\begin{tabular}{|c||r|r|r|r|r|r|r|} \hline
\rs\ [GeV] & \asrs & stat. & exp. & \hw & \ar & $\xmu=2.0$ & $\xmu=0.5$  \\ 
\hline
 14.0 & 0.1475 & $\pm$0.0014 & $\pm$0.0032 &   +0.0093 & +0.0117 & +0.0030 & $-$0.0018\\
 22.0 & 0.1442 & $\pm$0.0018 & $\pm$0.0028 & $-$0.0005 & +0.0038 & +0.0018 & $-$0.0002\\
 34.6 & 0.1358 & $\pm$0.0007 & $\pm$0.0018 & $-$0.0031 & +0.0010 & +0.0011 &   +0.0007\\
 35.0 & 0.1405 & $\pm$0.0006 & $\pm$0.0017 & $-$0.0033 & +0.0009 & +0.0012 &   +0.0008\\
 38.3 & 0.1366 & $\pm$0.0021 & $\pm$0.0045 & $-$0.0038 & +0.0009 & +0.0004 &   +0.0021\\
 43.8 & 0.1303 & $\pm$0.0012 & $\pm$0.0011 & $-$0.0038 & +0.0005 & +0.0001 &   +0.0020\\
\hline
\end{tabular}
\end{center}
\caption{The value of \as\ for each energy point and the statistical,
experimental, hadronization and scale errors.}
\label{fitresults}
\end{table}

\clearpage

\section*{ Figures }

\begin{figure}[htb!]
\begin{center}
\begin{tabular}{cc}
\includegraphics[width=0.4\textwidth]{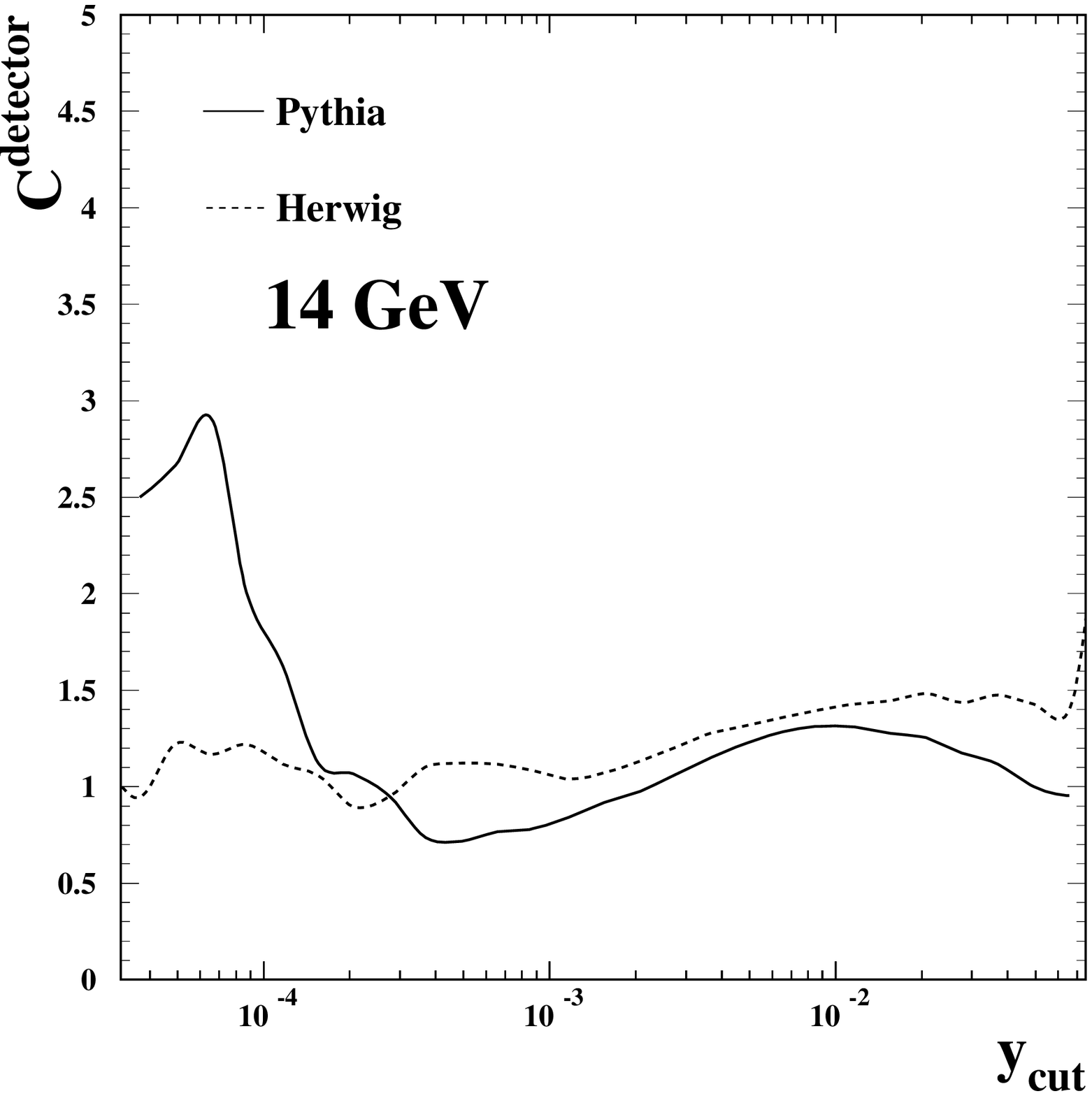} &
\includegraphics[width=0.4\textwidth]{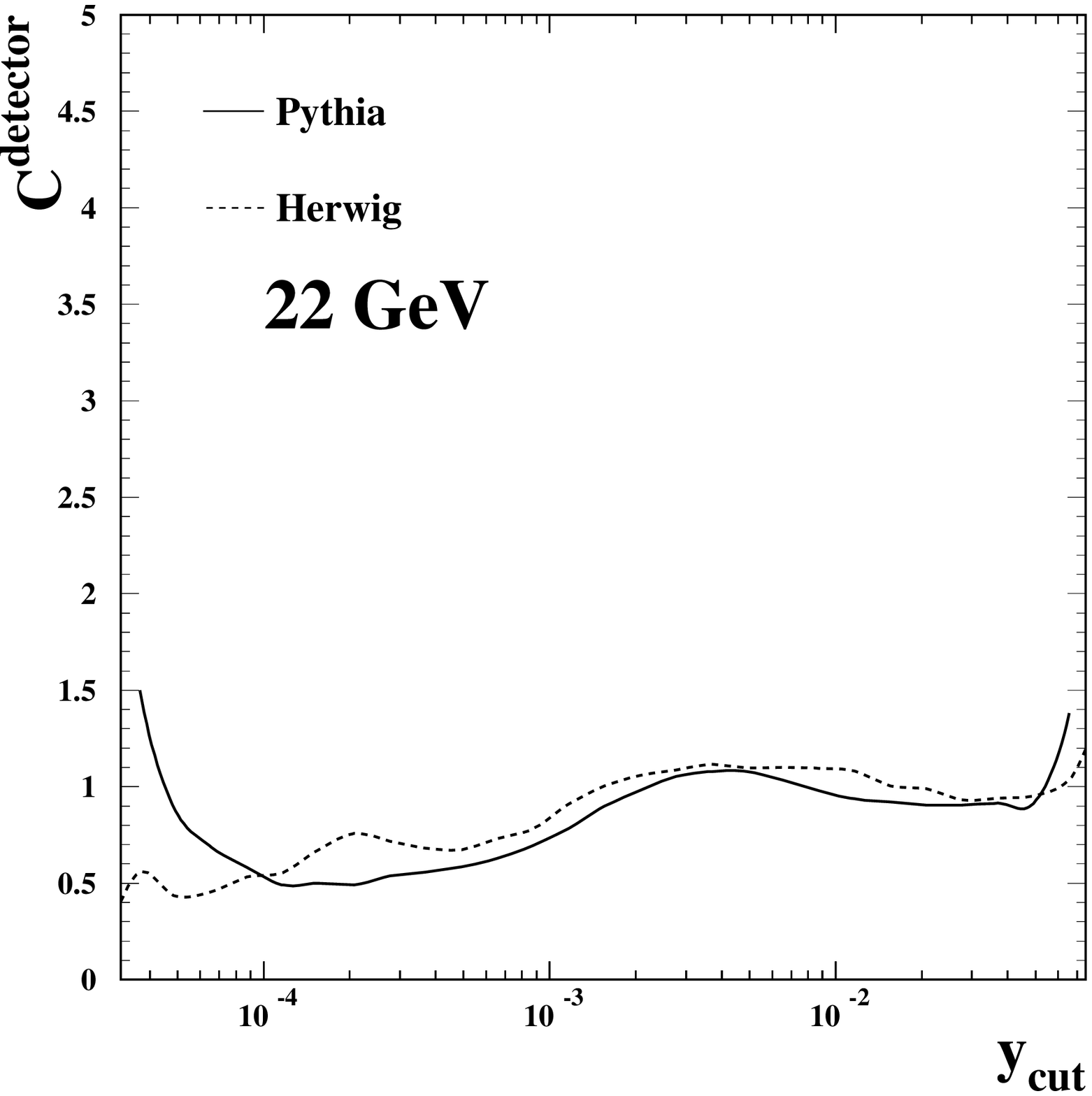} \\
\includegraphics[width=0.4\textwidth]{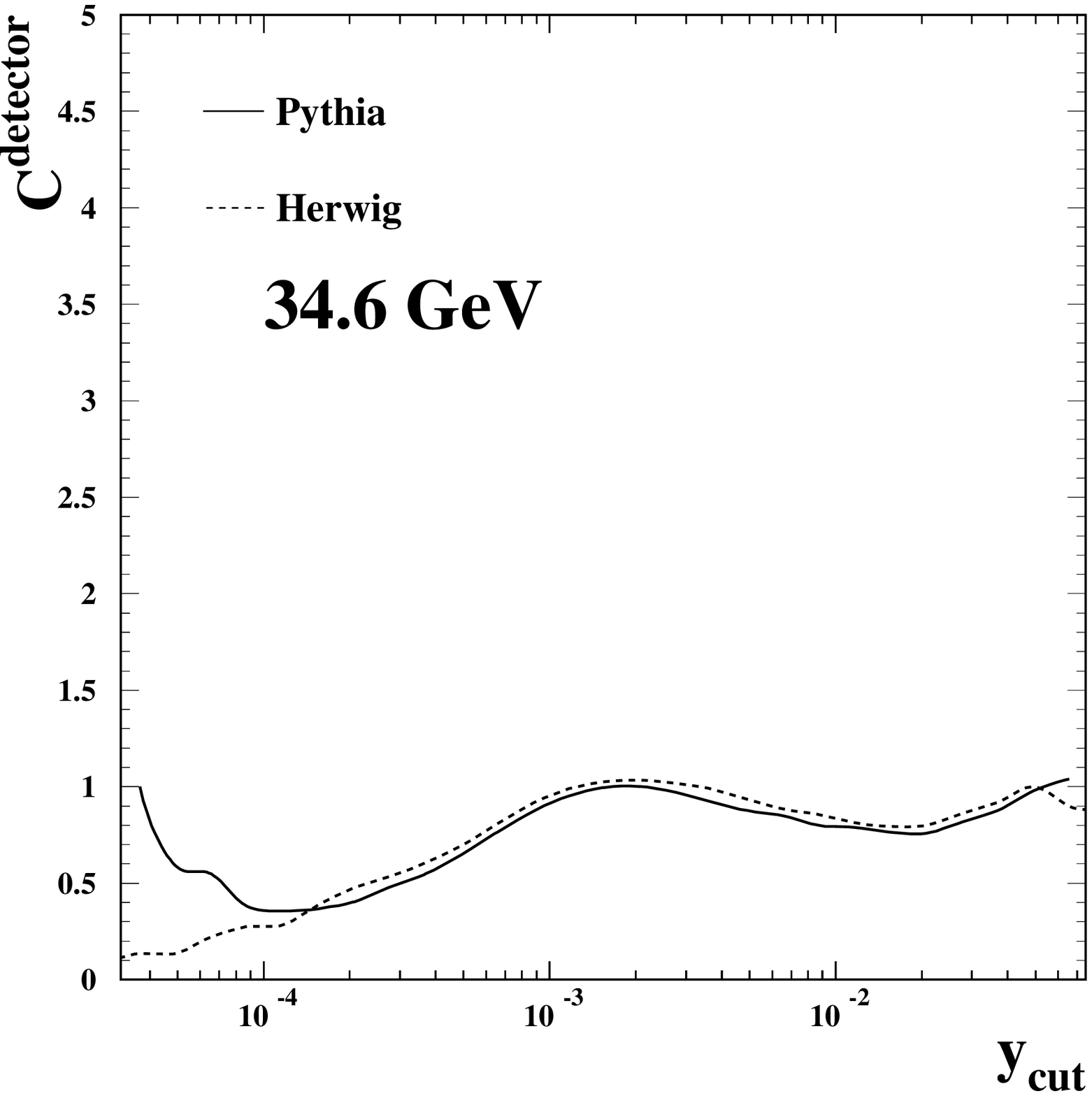} &
\includegraphics[width=0.4\textwidth]{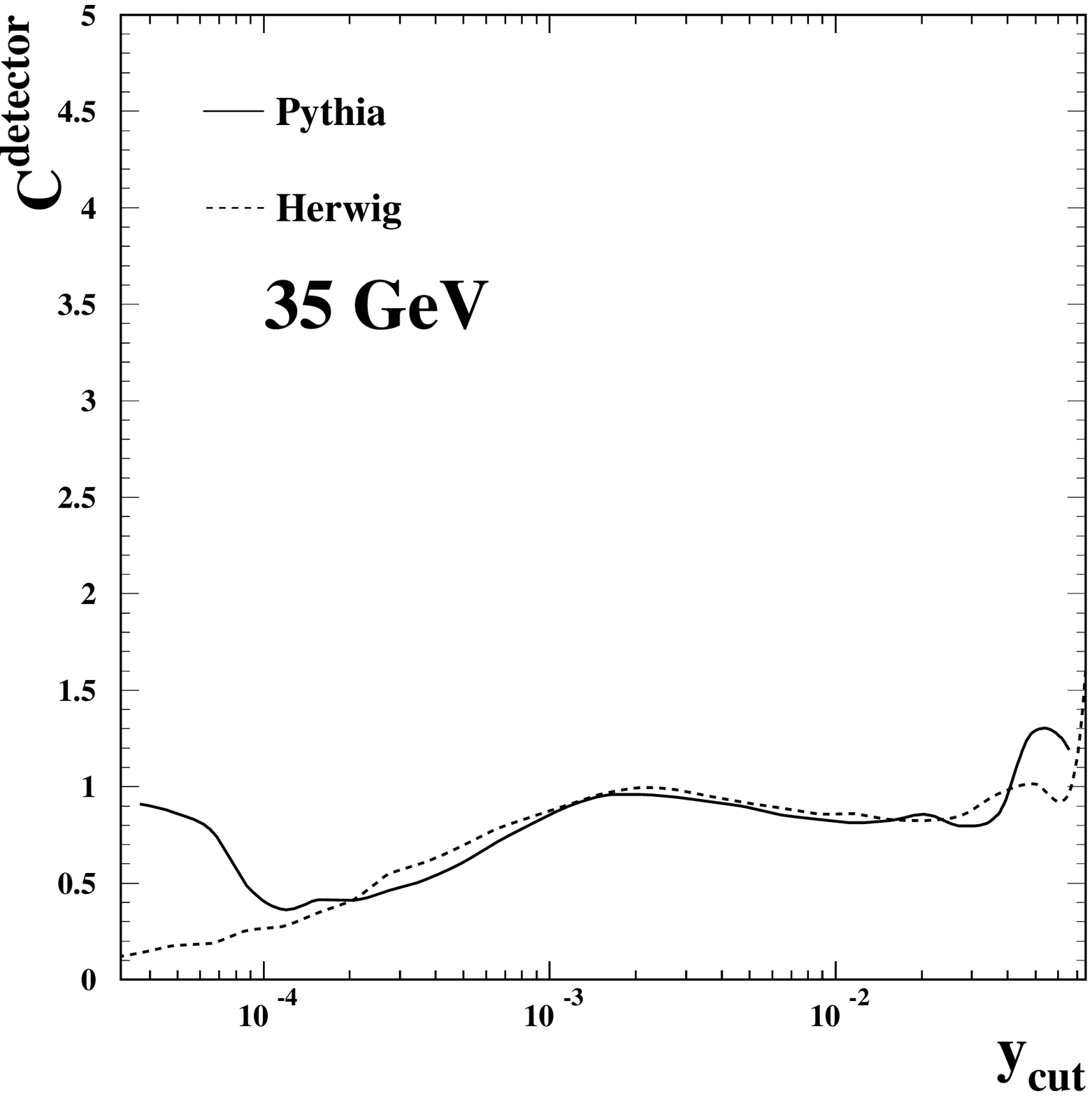} \\
\includegraphics[width=0.4\textwidth]{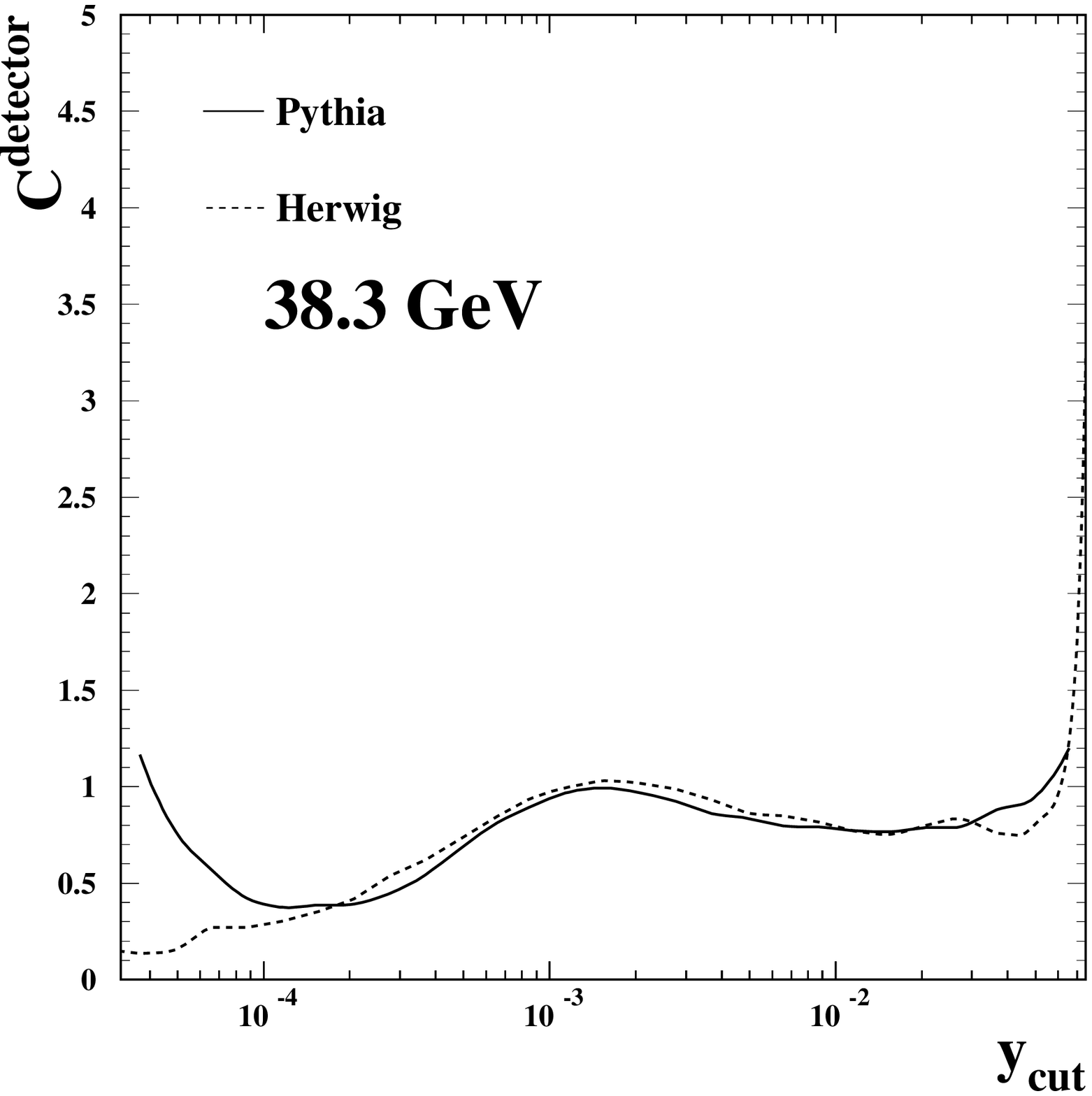} &
\includegraphics[width=0.4\textwidth]{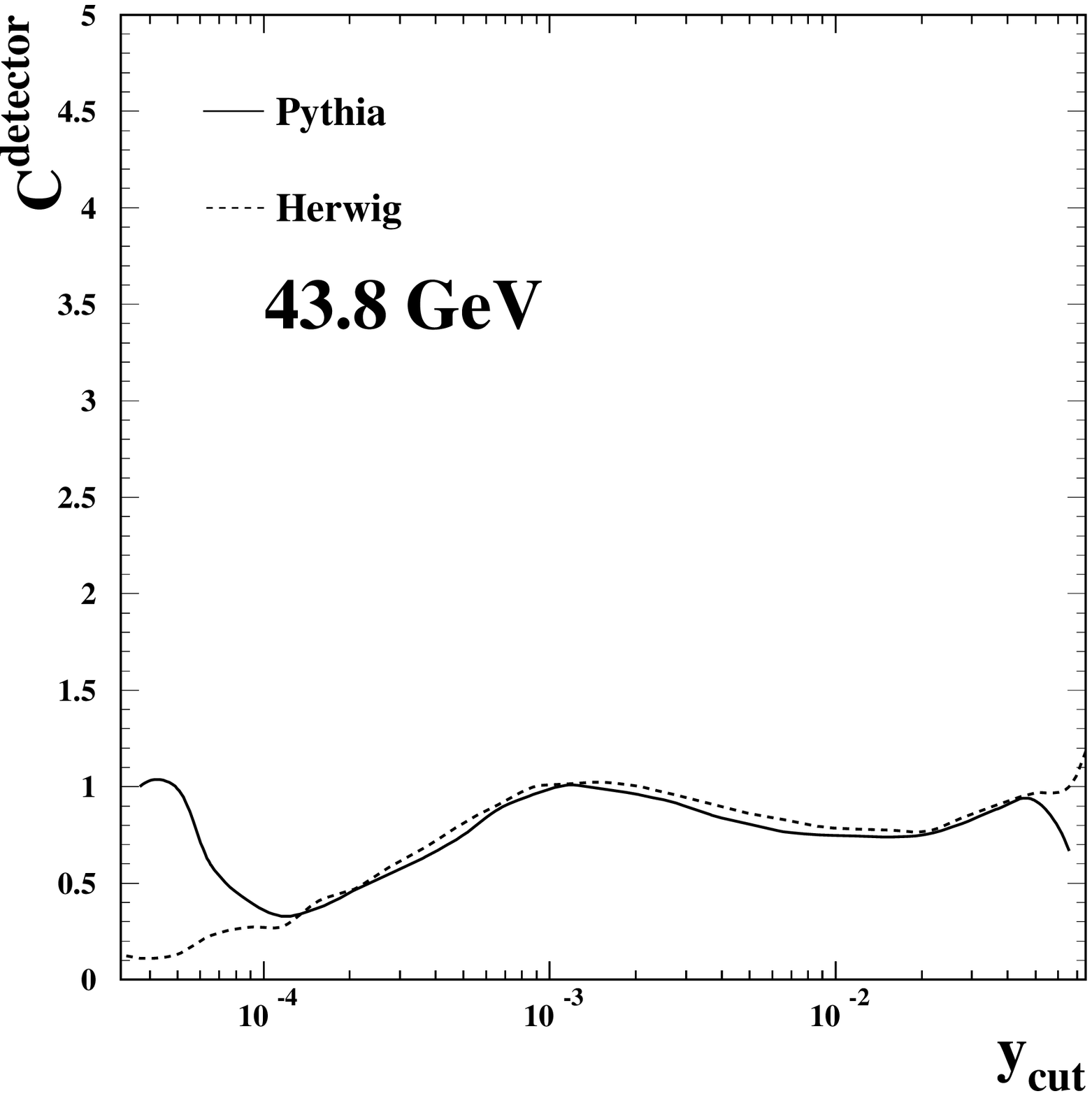} \\
\end{tabular}
\end{center}
\caption{The figures show the detector corrections for the four-jet
rate at the six energy points as calculated using \py\ and \hw.}
\label{detcor}
\end{figure}

\begin{figure}[htb!]
\begin{tabular}{cc}
\includegraphics[width=0.5\textwidth]{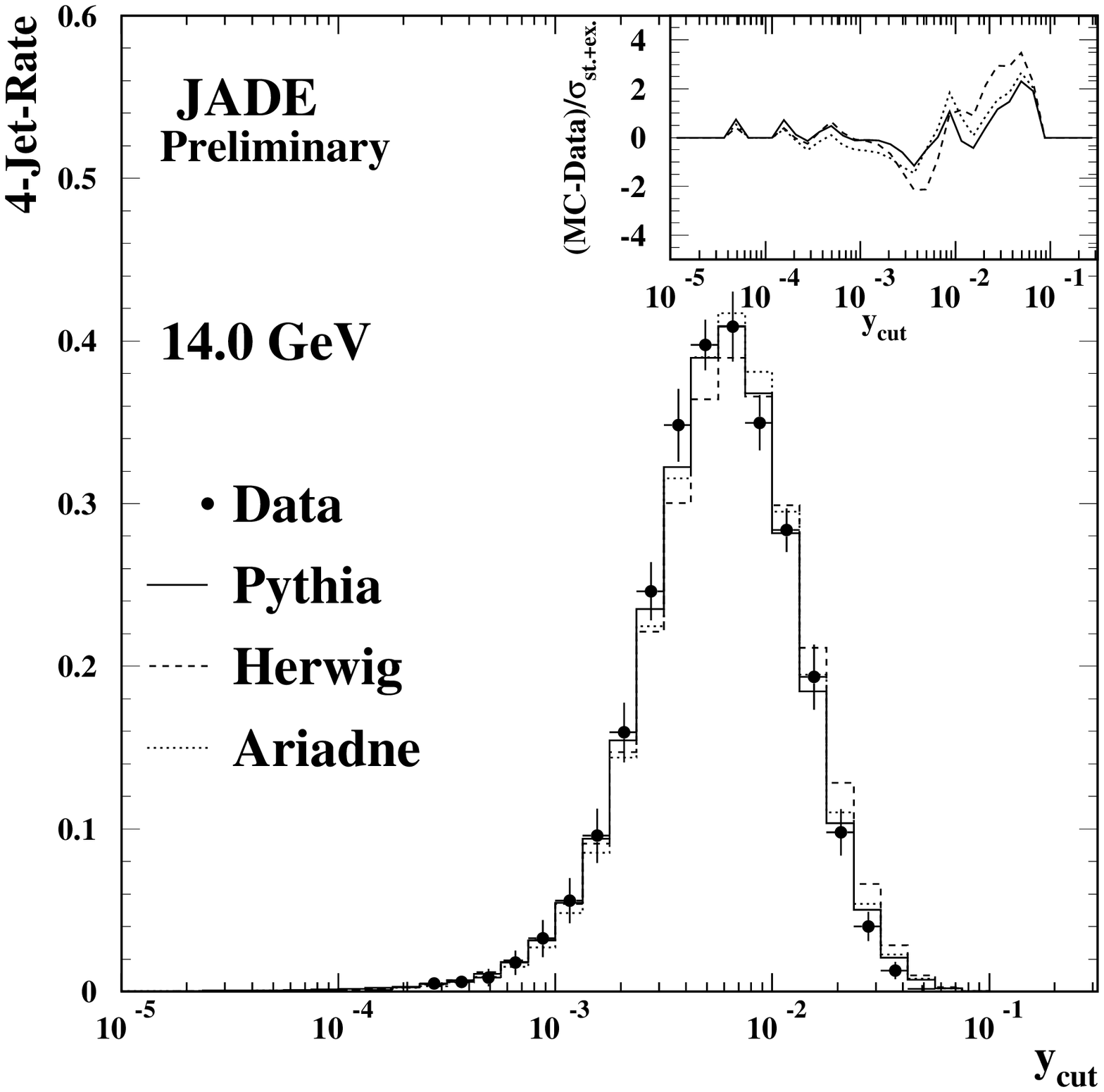} &
\includegraphics[width=0.5\textwidth]{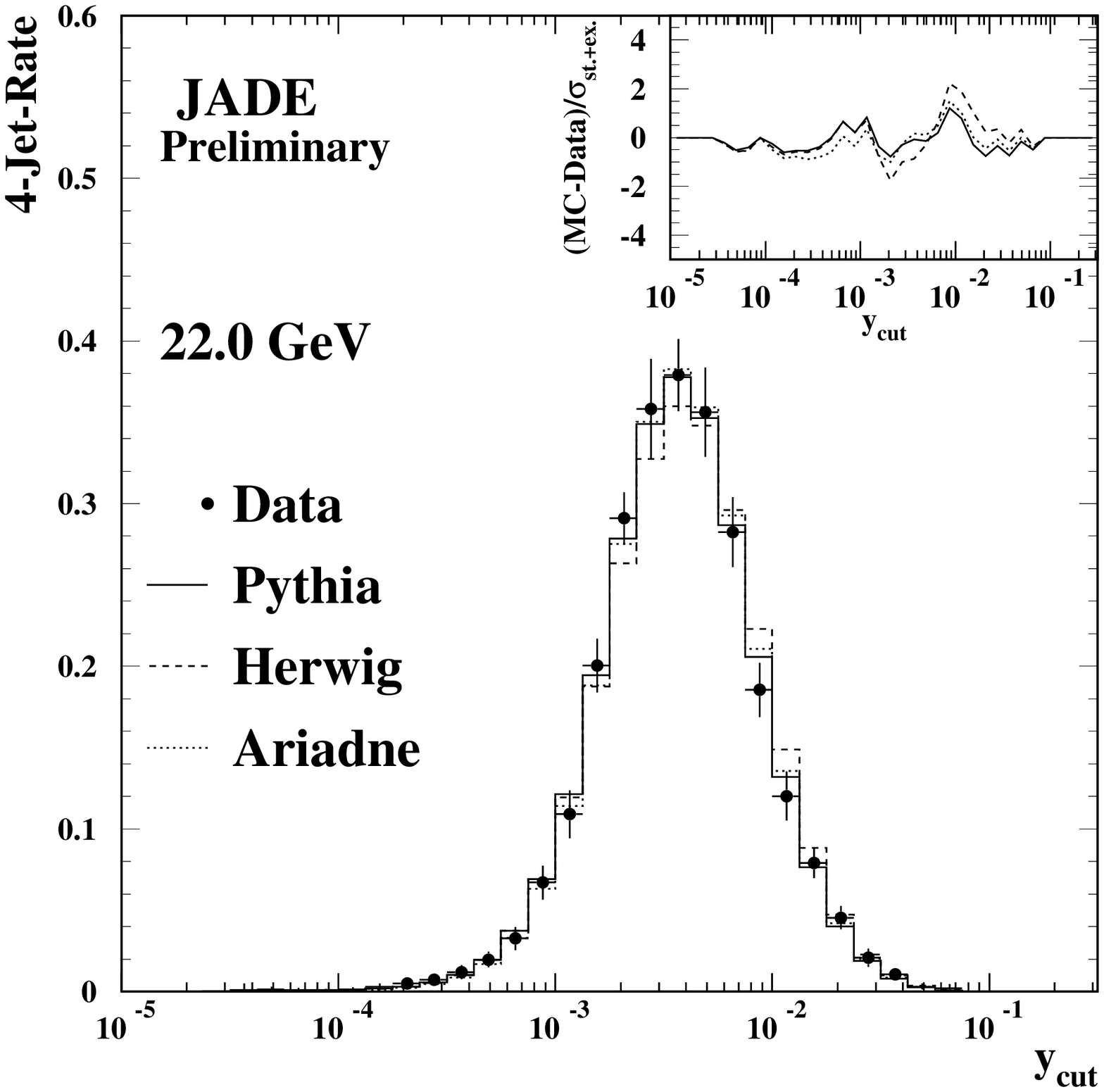} \\
\includegraphics[width=0.5\textwidth]{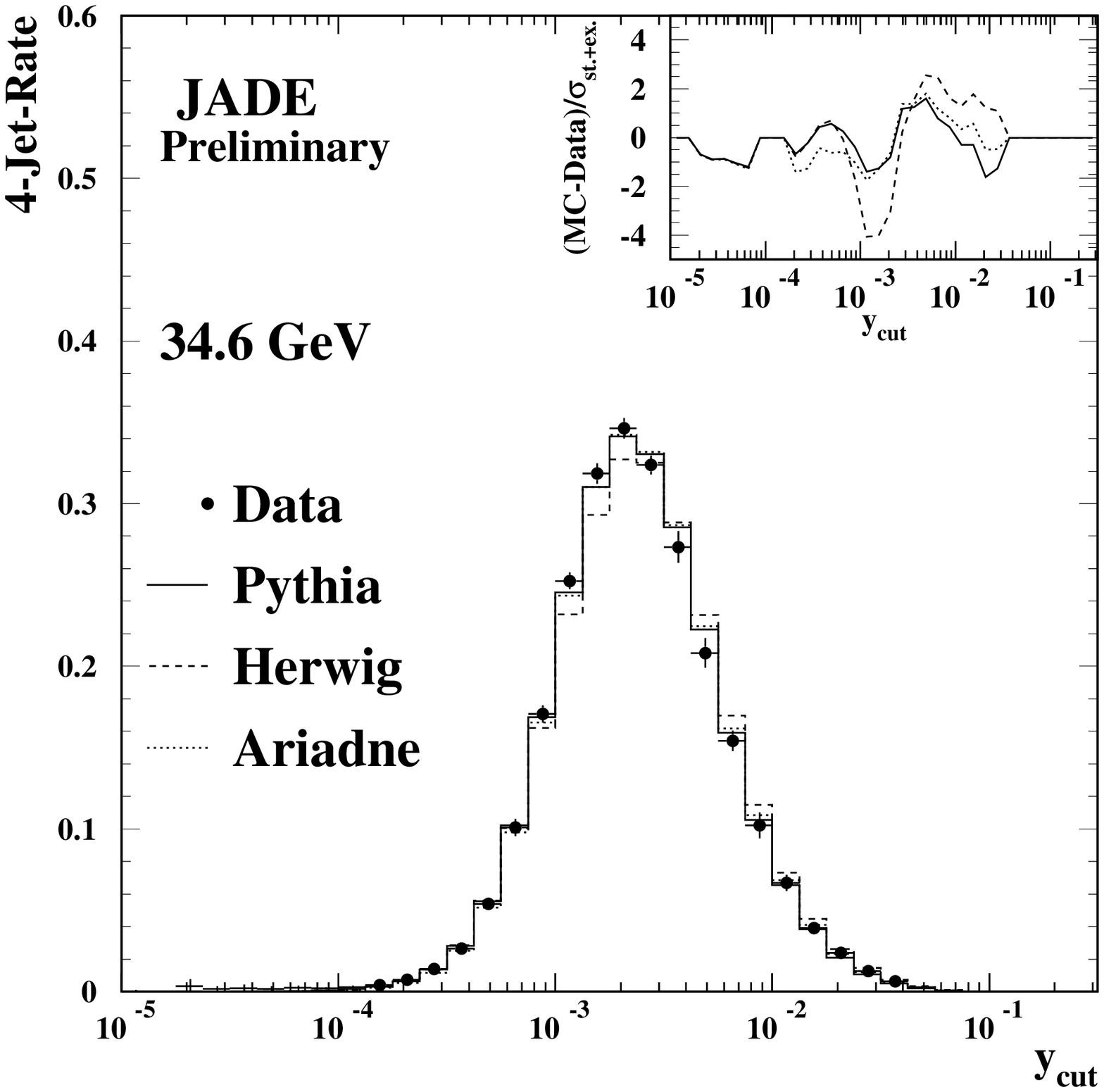} &
\includegraphics[width=0.5\textwidth]{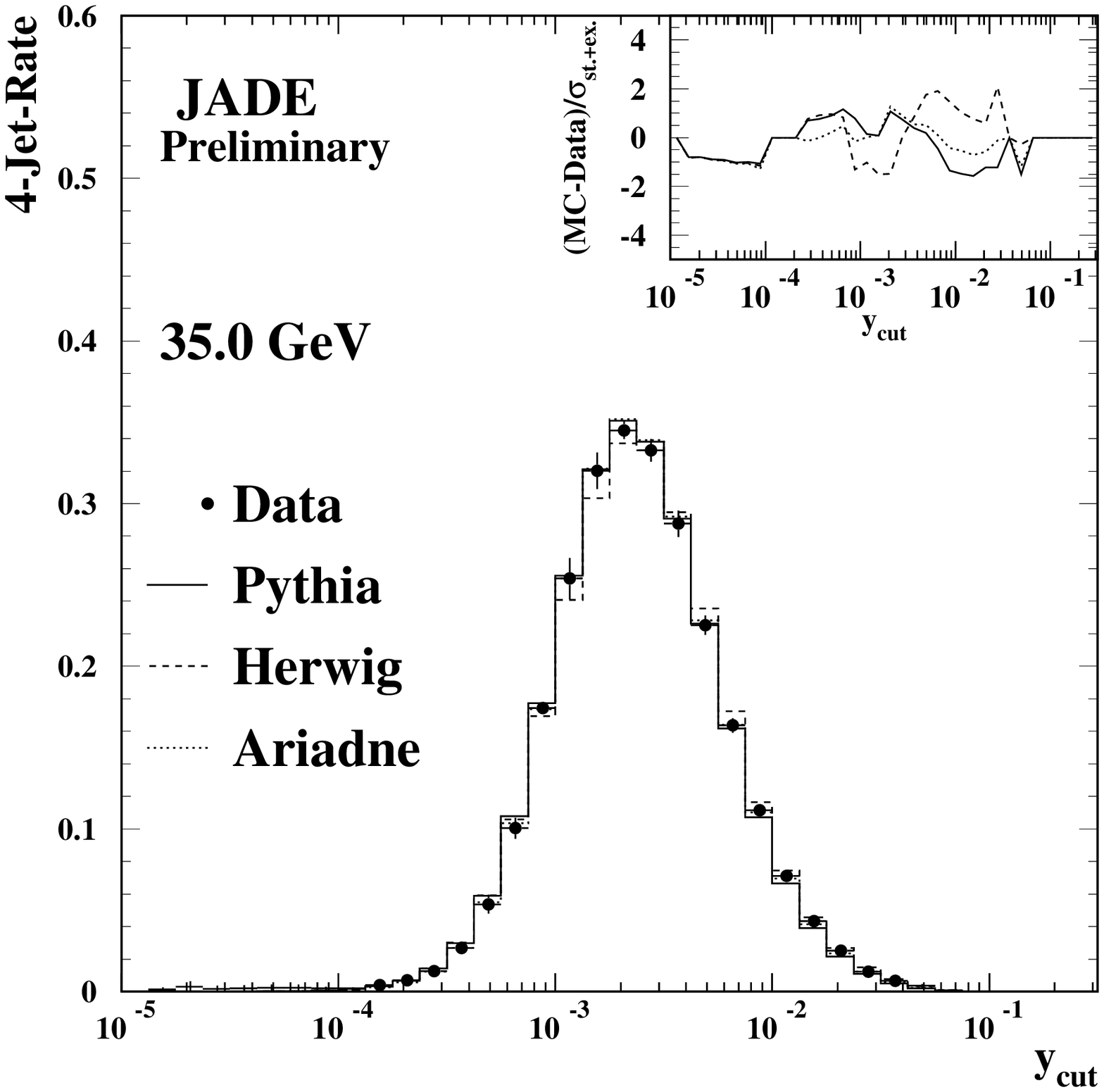} \\
\end{tabular}
\caption{The figures show the four-jet rate distribution
  at hadron level as a function of the \ycut\ resolution
  parameter obtained with the Durham algorithm.  The four-jet rate at
  five average centre-of-mass energies for the data, corrected to the
  hadron level, are shown for $\rs=14$ to 35~GeV in comparison with
  predictions based on \py, \hw\ and \ar\ Monte Carlo events.  The
  errors shown include all statistical and experimental uncertainties.
  The panel in each upper right corner shows the differences between
  data and Monte Carlo, divided by the sum of the statistical and
  experimental error. At points with no data events, the difference is
  set to zero. }
\label{hadron}
\end{figure}

\begin{figure}[htb!]
\begin{tabular}{cc}
\includegraphics[width=0.5\textwidth]{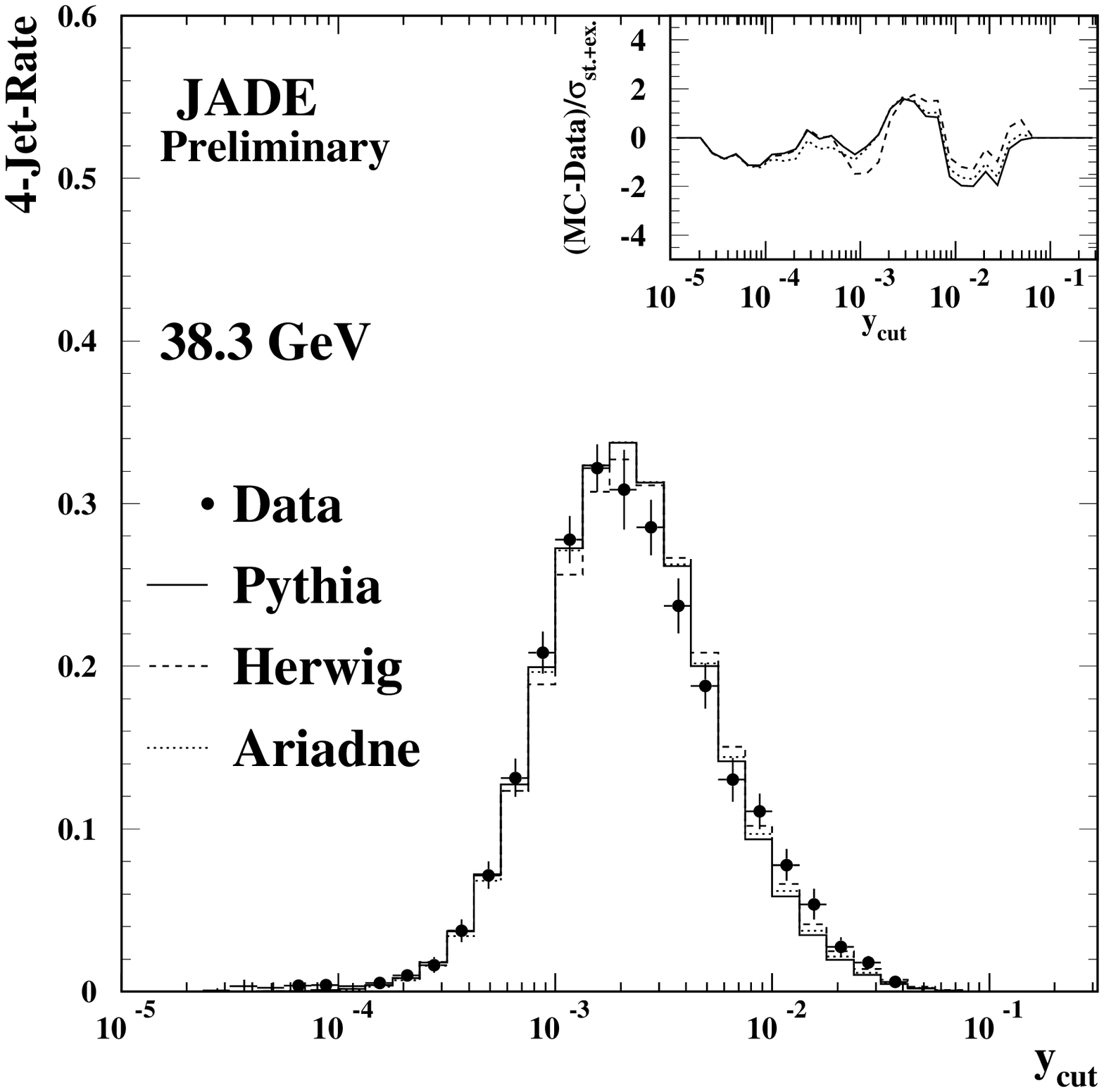} &
\includegraphics[width=0.5\textwidth]{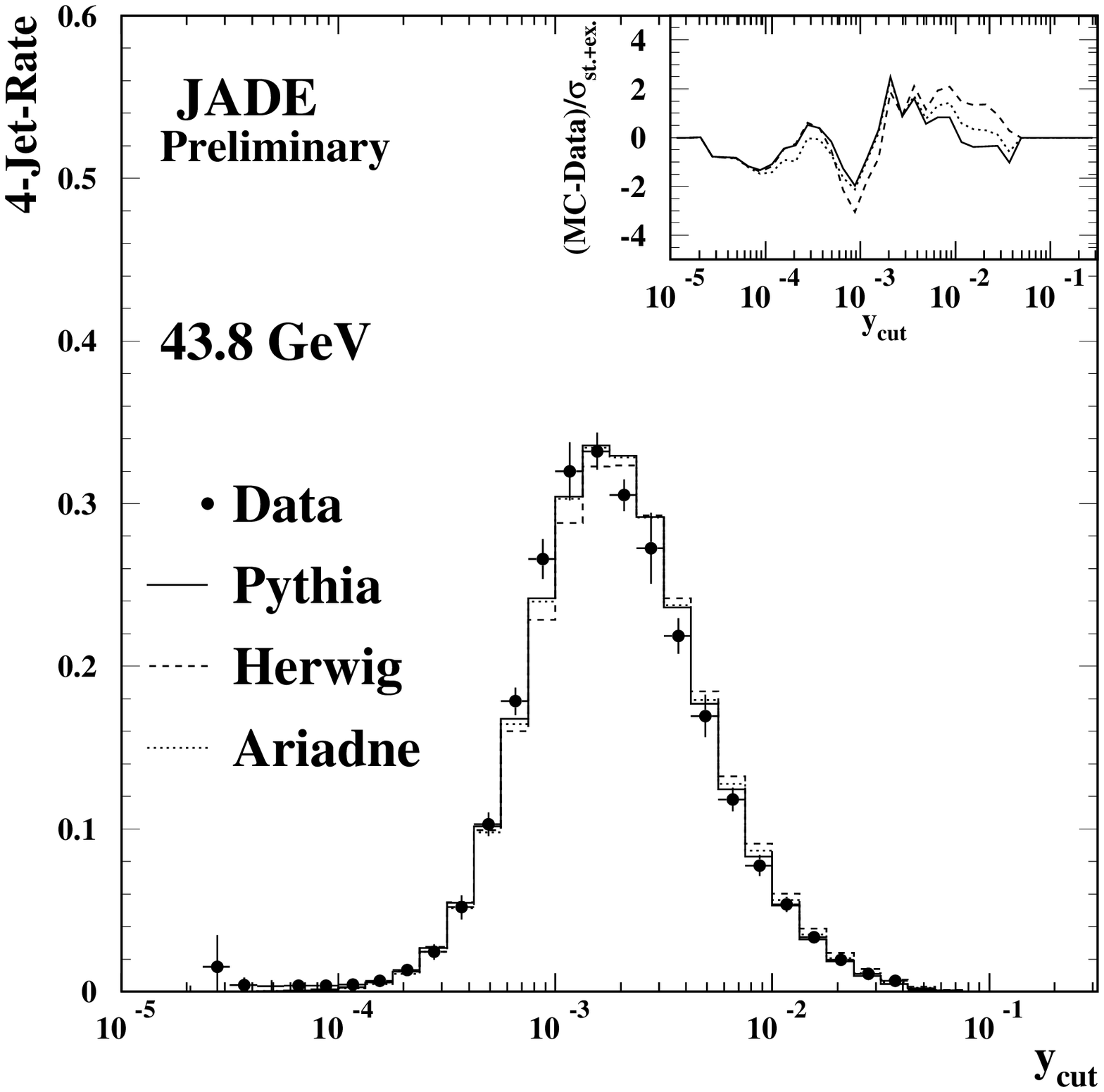} \\
\end{tabular}
\caption{ Same as figure~\ref{hadron} for $\rs=38.3$ and 43.8~GeV. }
\label{hadron2}
\end{figure}

\begin{figure}[htb!]
\begin{center}
\begin{tabular}{cc}
\includegraphics[width=0.4\textwidth]{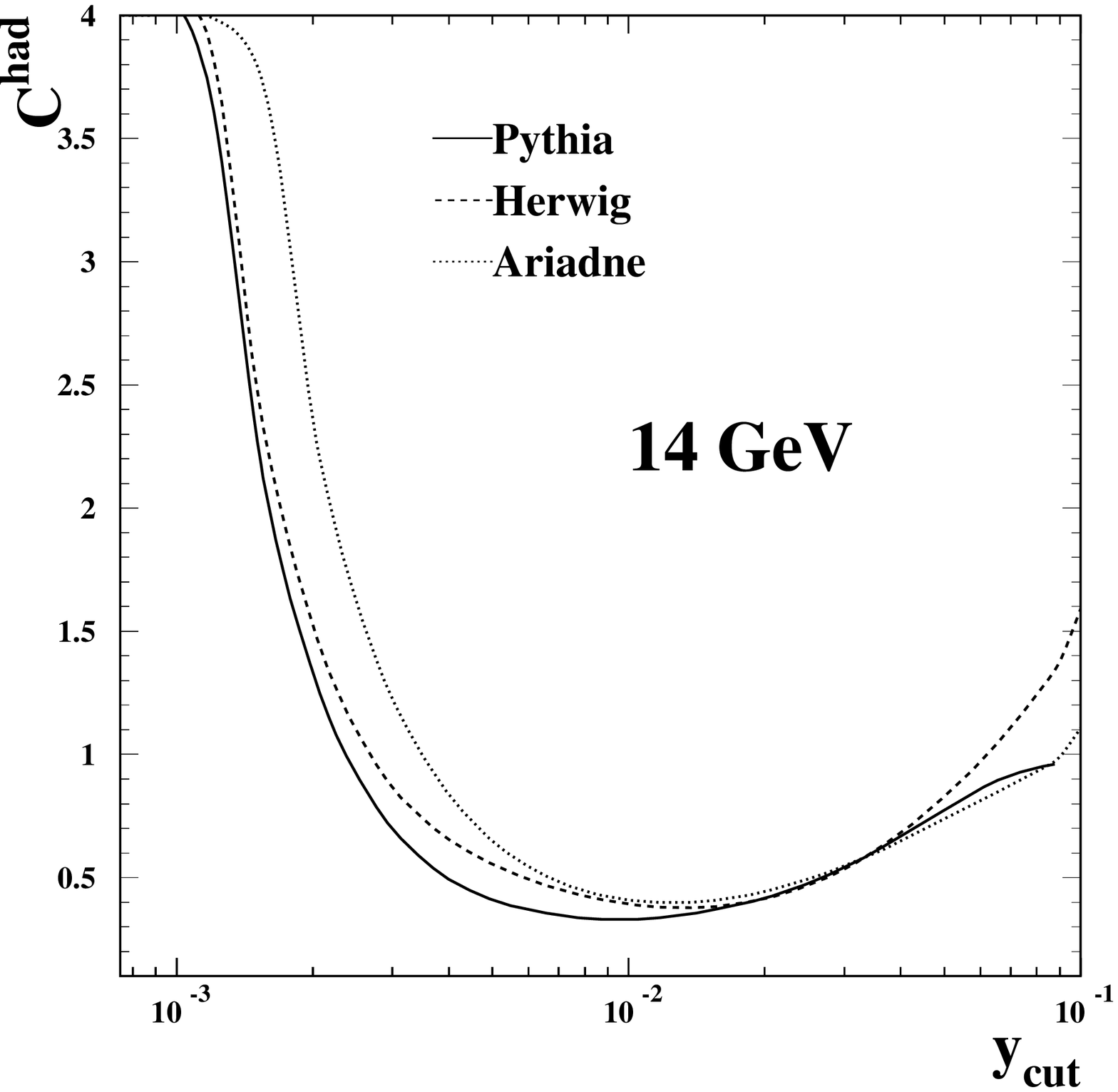} &
\includegraphics[width=0.4\textwidth]{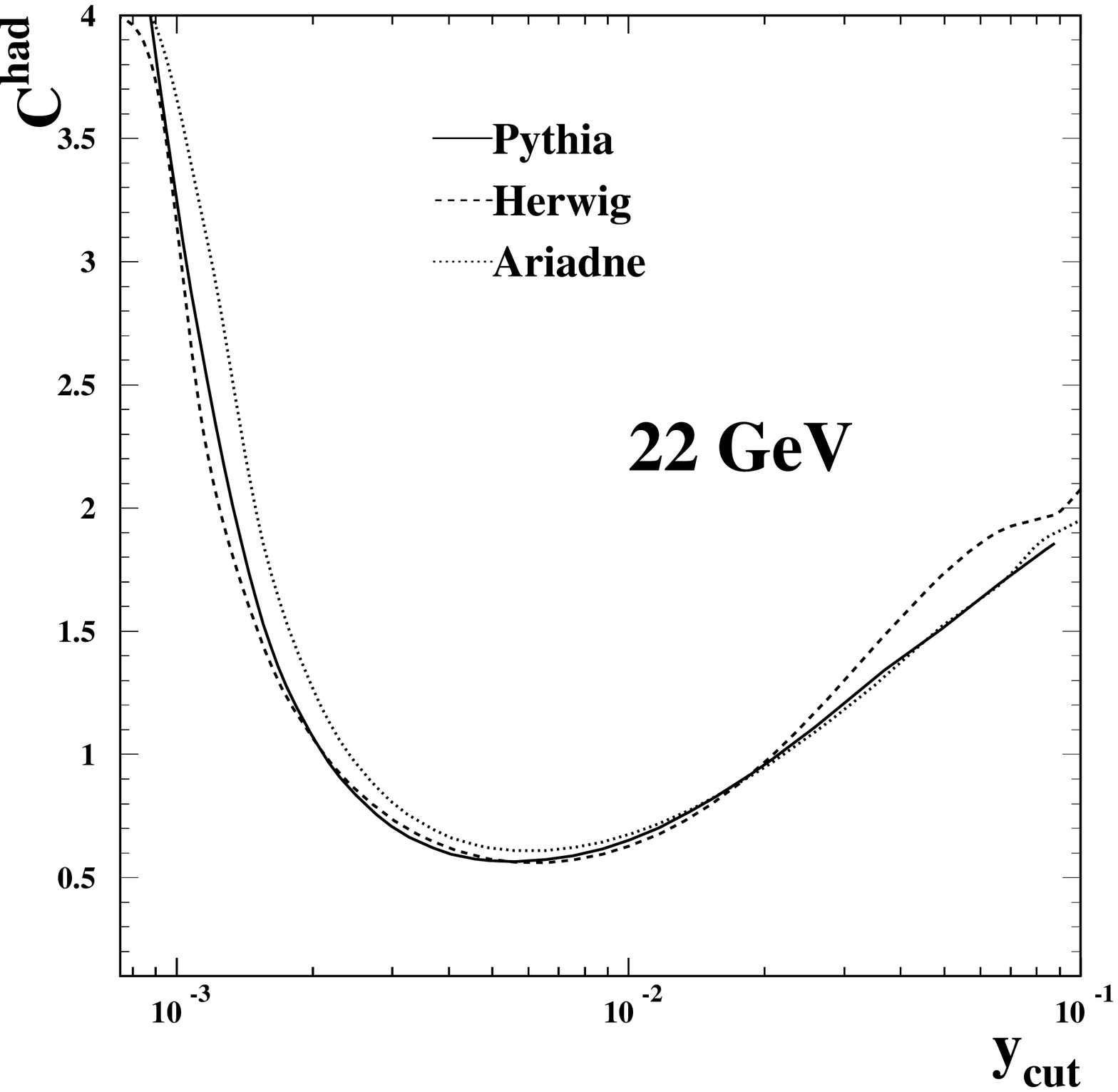} \\
\includegraphics[width=0.4\textwidth]{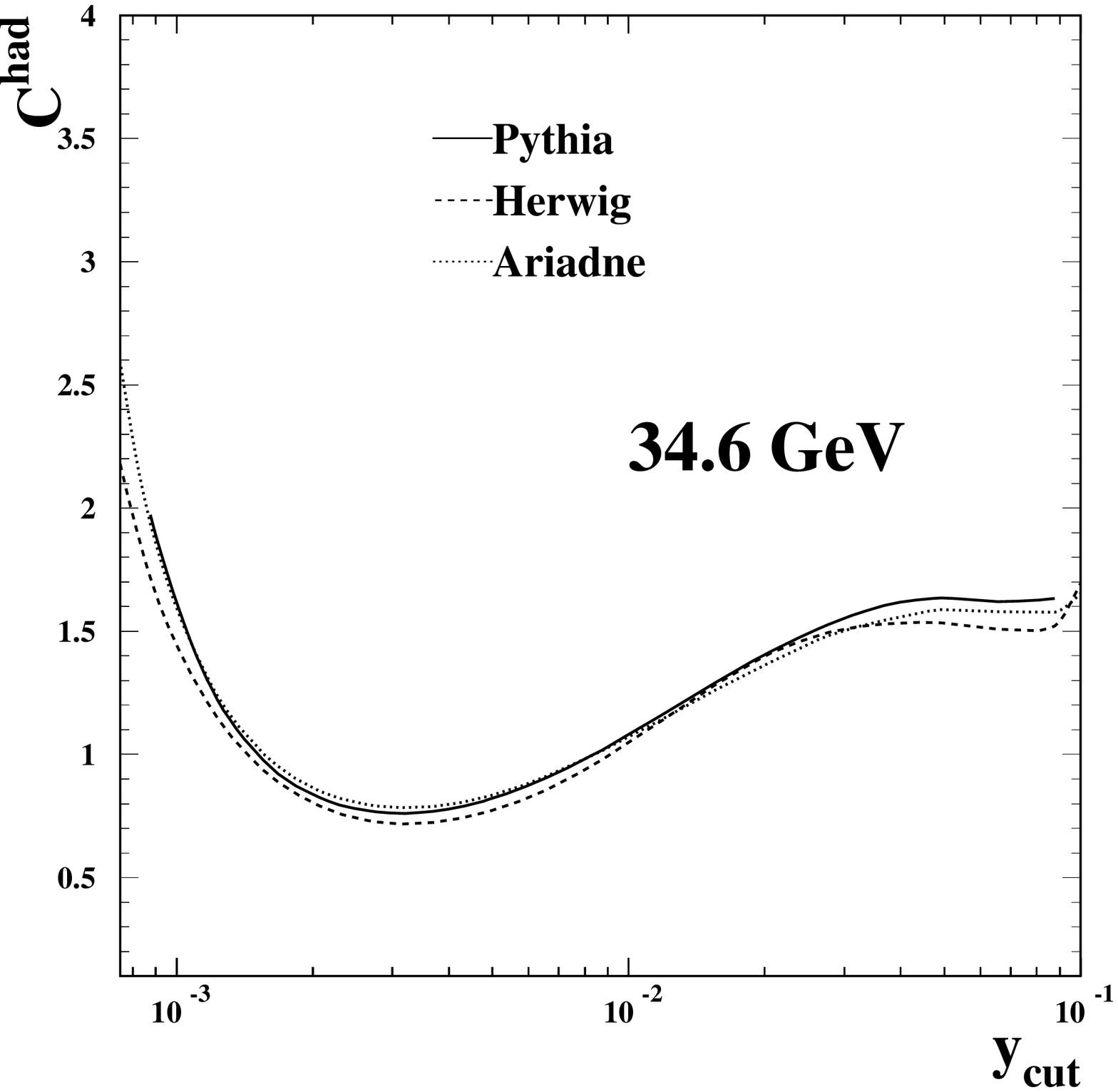} &
\includegraphics[width=0.4\textwidth]{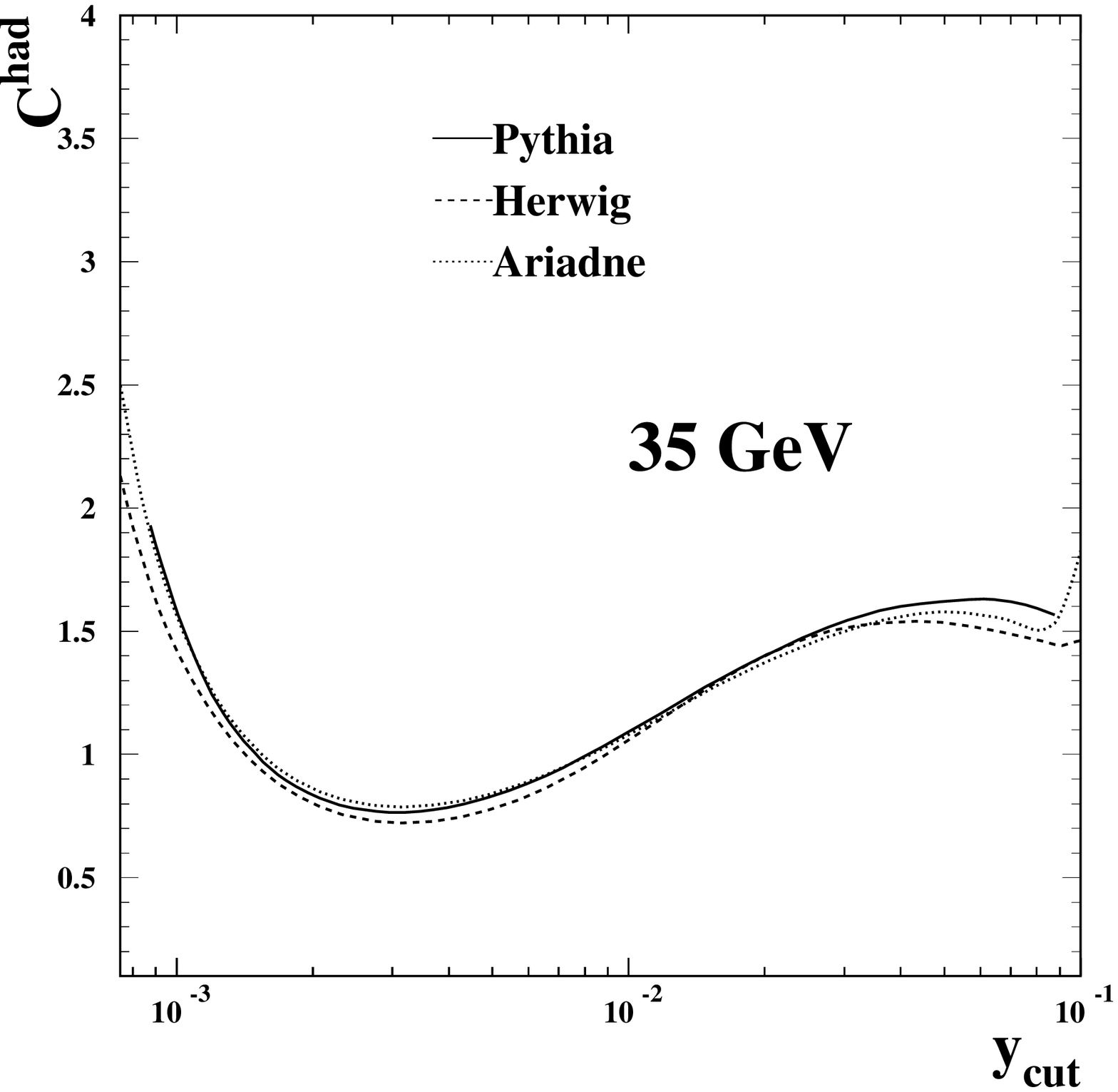} \\
\includegraphics[width=0.4\textwidth]{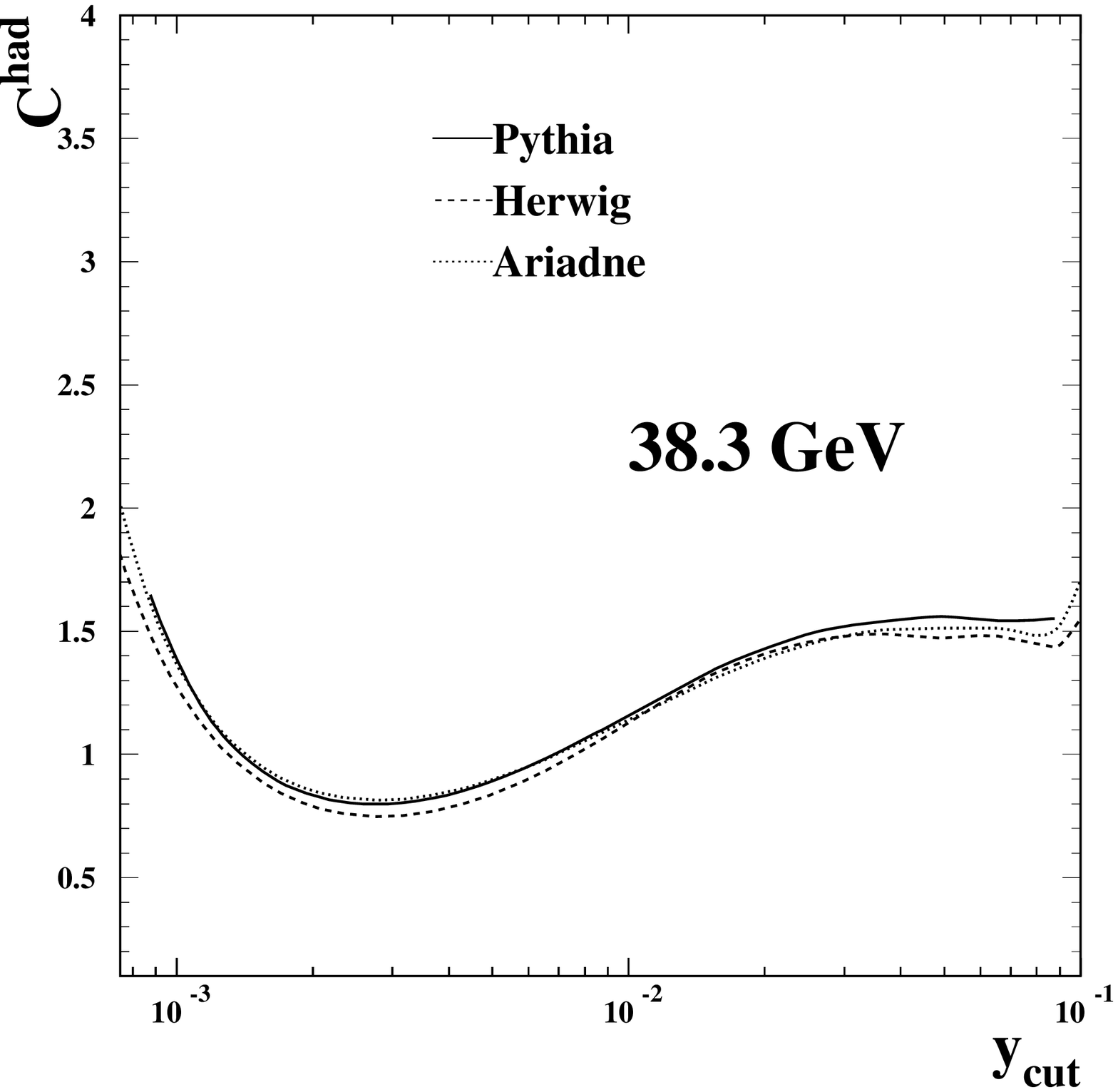} &
\includegraphics[width=0.4\textwidth]{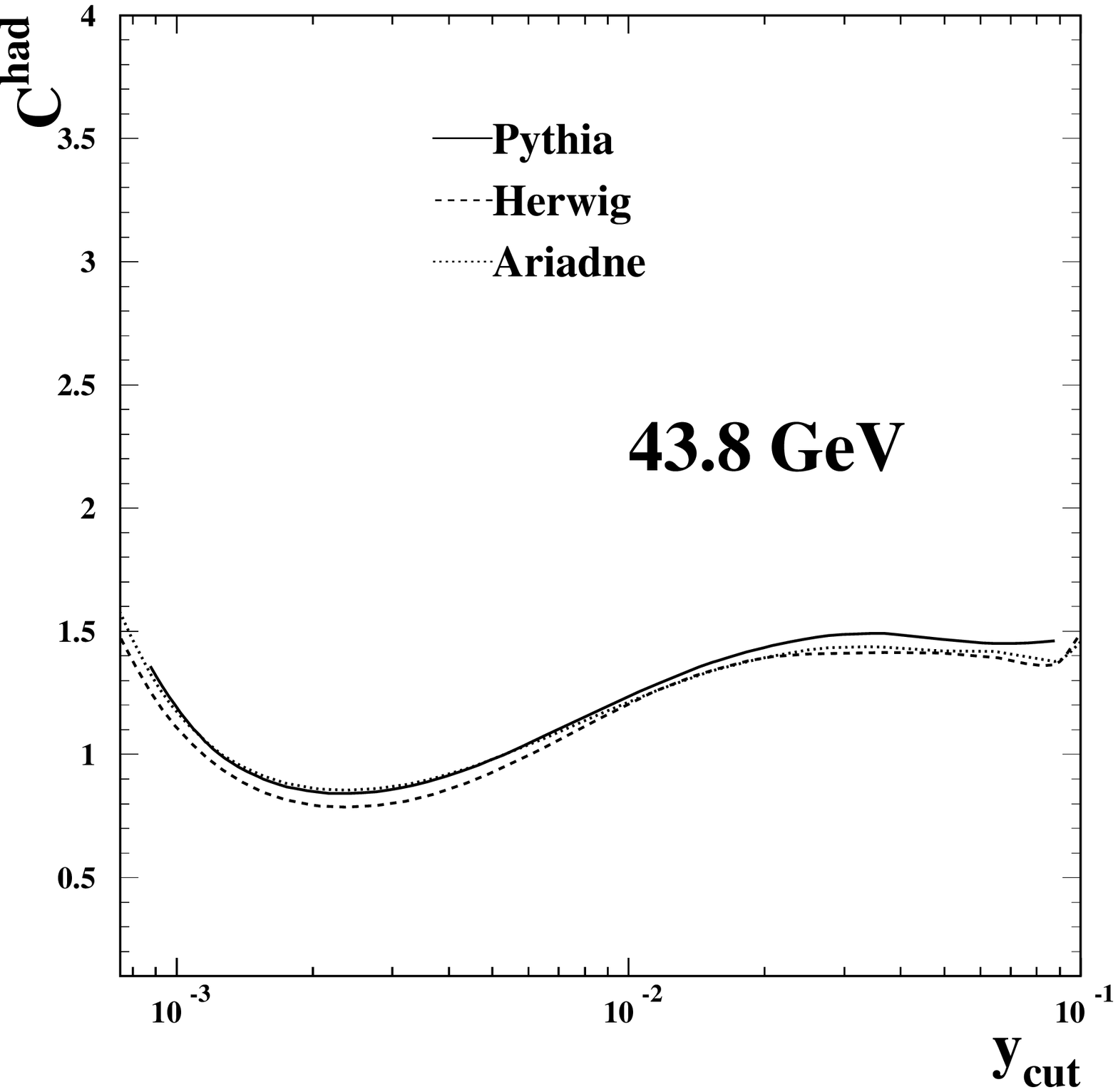} \\
\end{tabular}
\end{center}
\caption{The figures show the hadronization corrections for the four-jet
rate at the six energy points as calculated using \py, \hw\ and \ar.}
\label{hadcor}
\end{figure}

\begin{figure}[htb!]
\begin{tabular}{cc}
\includegraphics[width=0.5\textwidth]{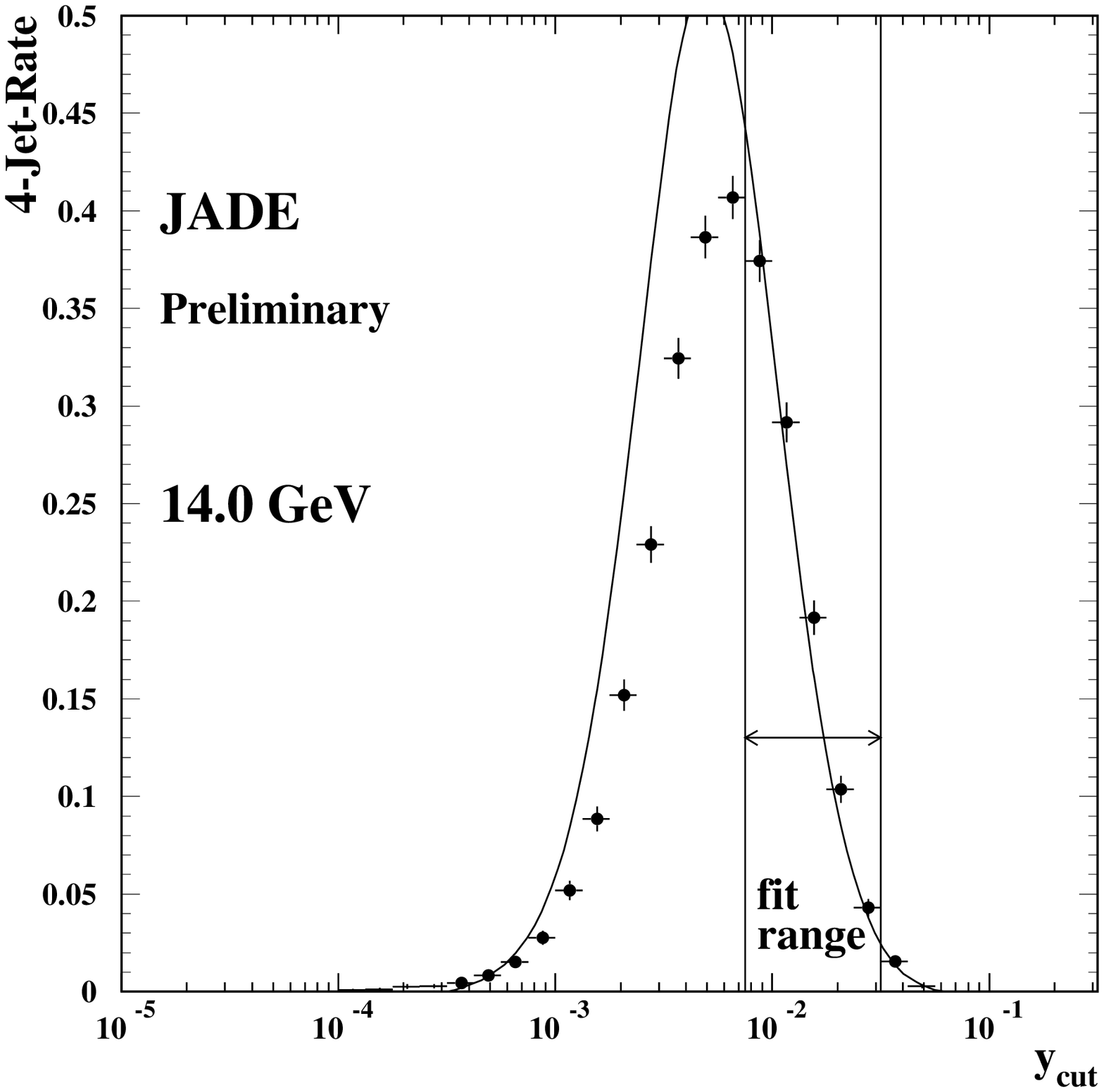} &
\includegraphics[width=0.5\textwidth]{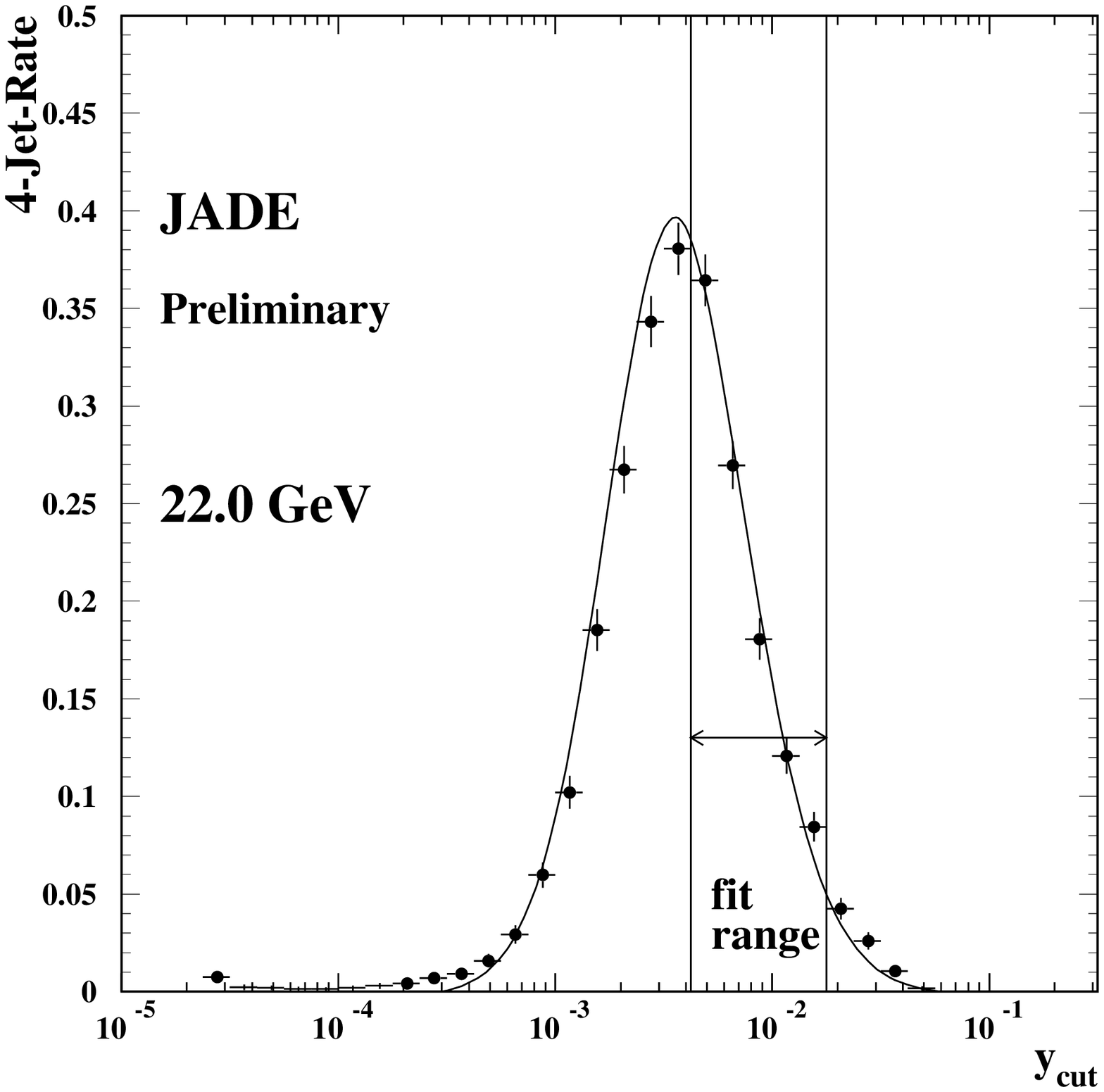} \\
\includegraphics[width=0.5\textwidth]{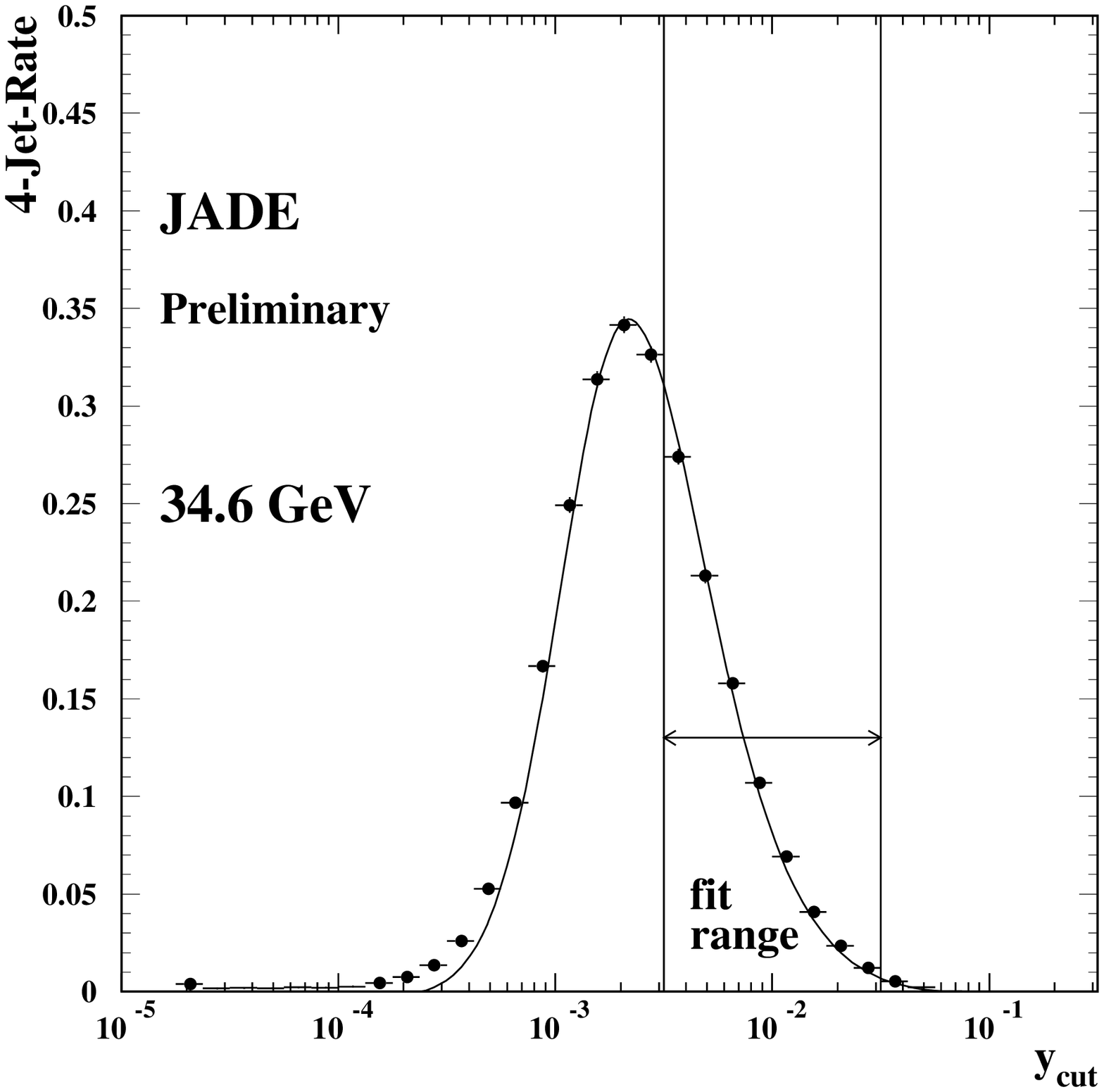} &
\includegraphics[width=0.5\textwidth]{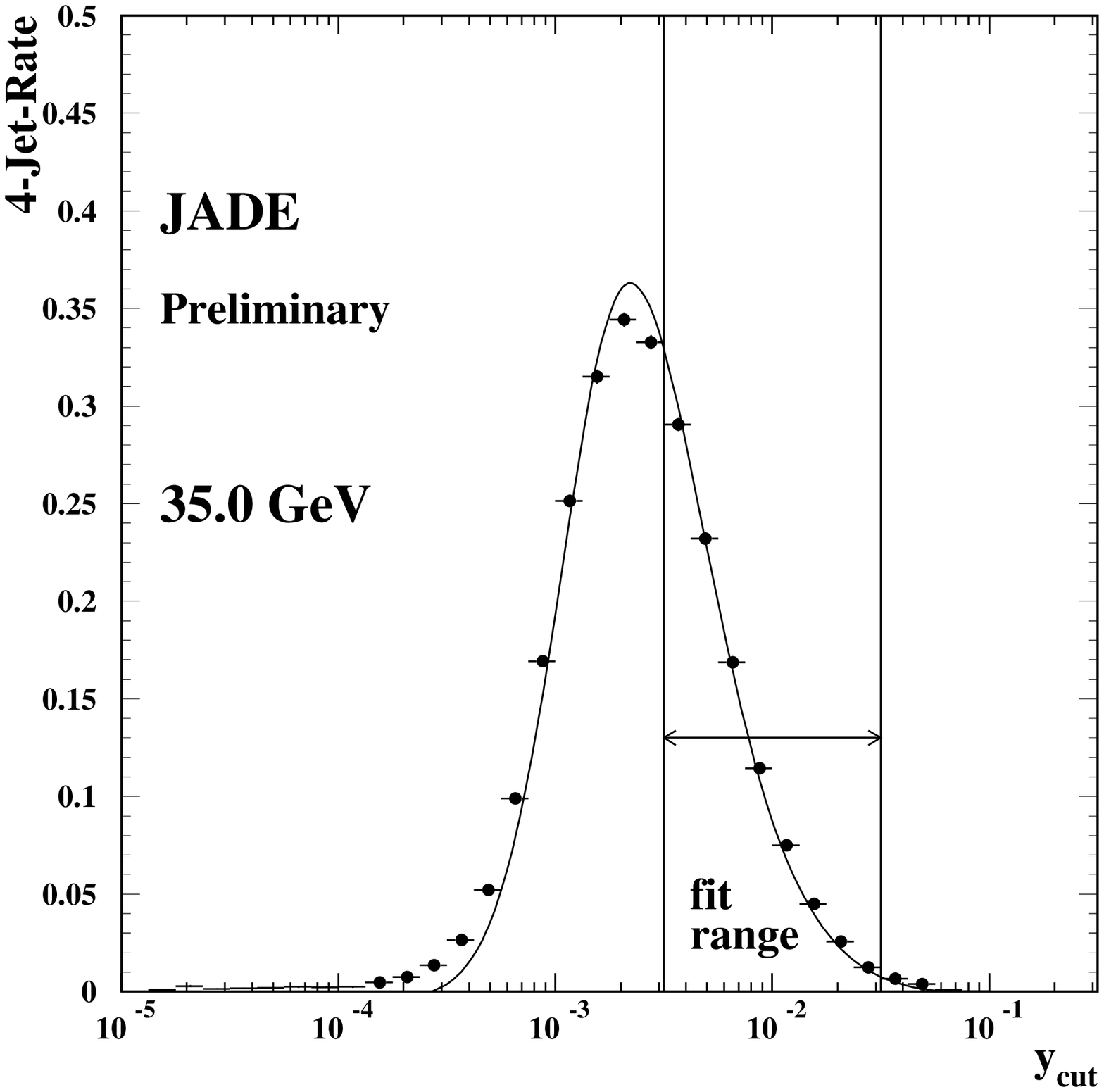} \\
\end{tabular}
\caption{ The plots show the four-jet rate distributions at the hadron level
  for $\rs=14$~GeV to 35~GeV. The errors correspond to the statistical
  error only and the curves indicate the theory prediction after using
  the fit results. }
\label{fit_plot}
\end{figure}

\begin{figure}[htb!]
\begin{tabular}{cc}
\includegraphics[width=0.5\textwidth]{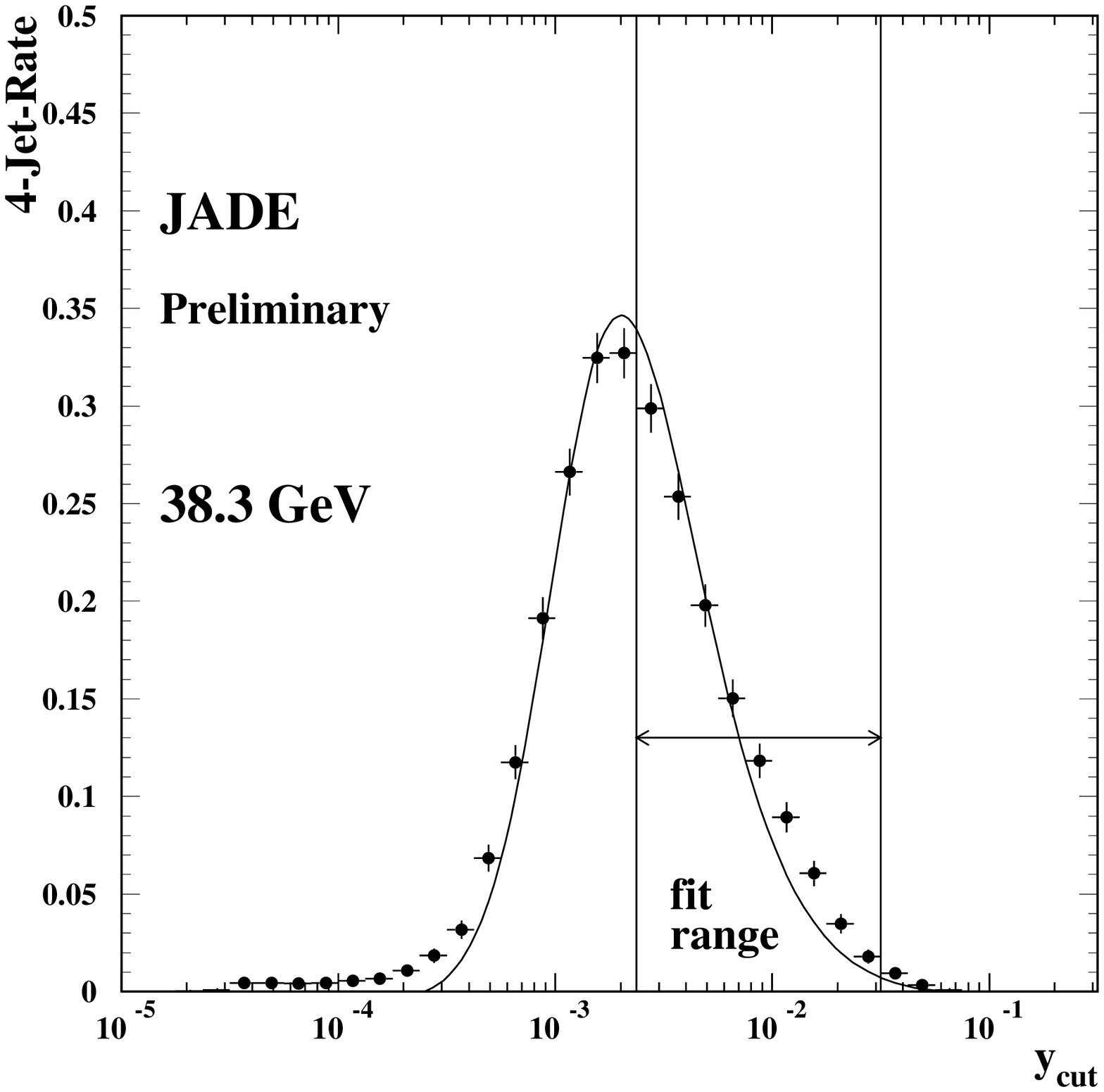} &
\includegraphics[width=0.5\textwidth]{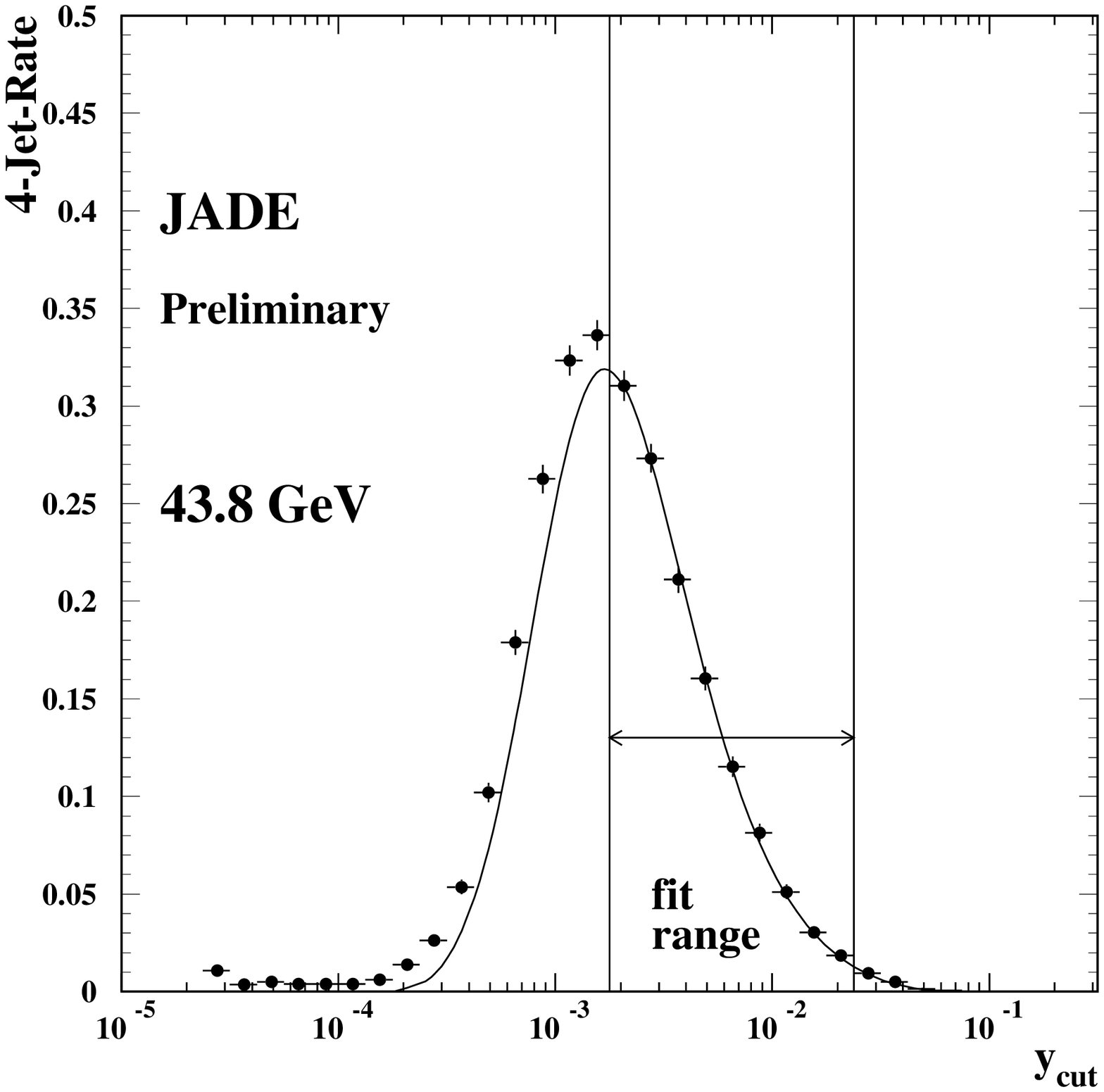} \\
\end{tabular}
\caption{ Same as figure~\ref{fit_plot} for $\rs=38.3$~GeV and 43.8~GeV }
\label{fit_plot2}
\end{figure}

\begin{figure}[htb!]
\begin{center}
\includegraphics[width=1.0\textwidth]{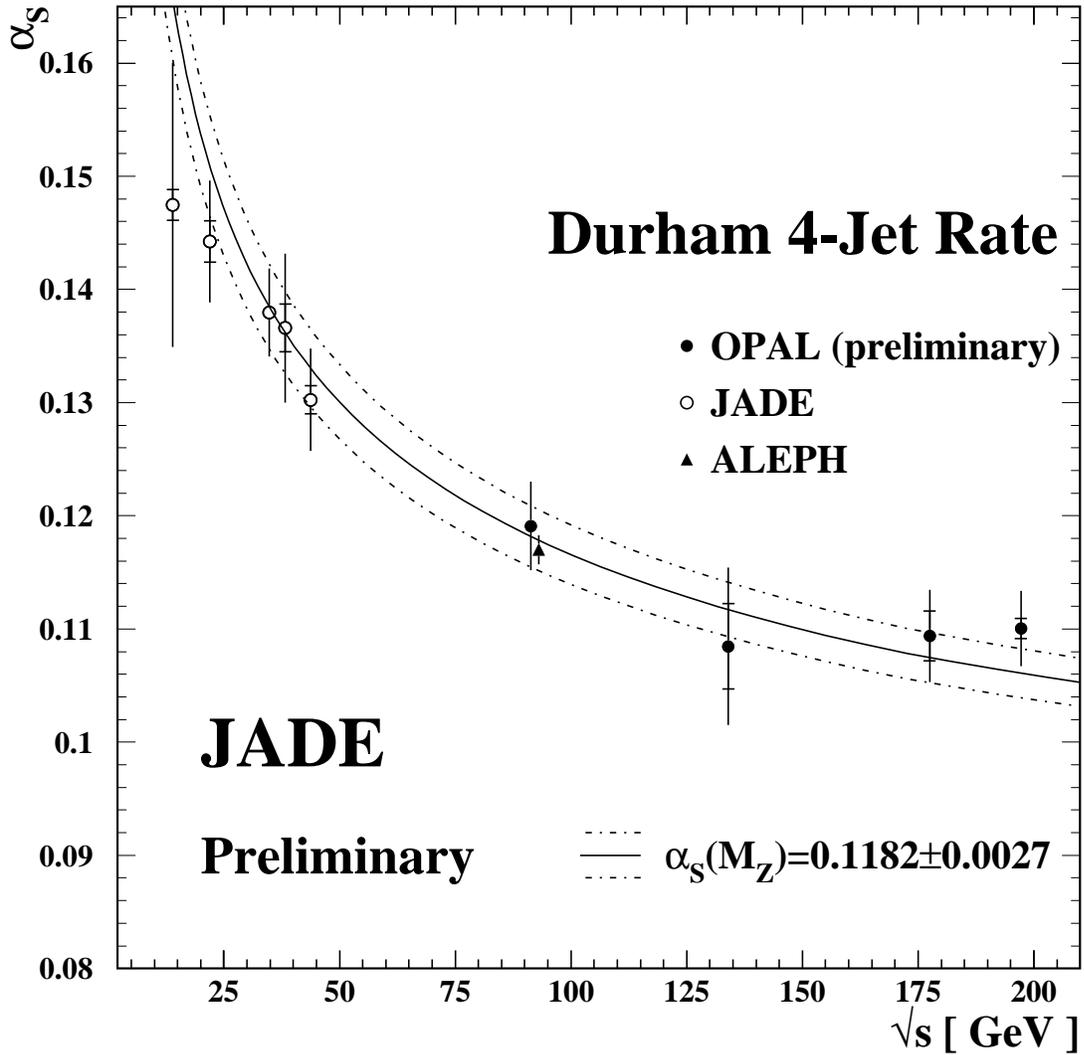}
\end{center}
\caption{The values for \as\ at the various energy points. The errors
  show the statistical (inner part) and the total error.  The full and
  dash-dotted lines indicate the current world average value of
  \asmz~\cite{bethke04}.  The results at $\rs=34.6$ and 35~GeV have
  been combined for clarity.  The results from ALEPH~\cite{aleph249}
  and OPAL~\cite{OPALPN527} (preliminary) are shown as well. }
\label{alphas_fit}
\end{figure}

\end{document}